\documentclass{aa}  
\usepackage{multirow}
\usepackage{natbib,twoopt}
\usepackage[utf8]{inputenc}

\usepackage[breaklinks=true]{hyperref} 
\bibpunct{(}{)}{;}{a}{}{,}             
\makeatletter
  \newcommandtwoopt{\citeads}[3][][]{\href{http://adsabs.harvard.edu/abs/#3}%
    {\def\hyper@linkstart##1##2{}%
     \let\hyper@linkend\@empty\citealp[#1][#2]{#3}}}
  \newcommandtwoopt{\citepads}[3][][]{\href{http://adsabs.harvard.edu/abs/#3}%
    {\def\hyper@linkstart##1##2{}%
     \let\hyper@linkend\@empty\citep[#1][#2]{#3}}}
  \newcommandtwoopt{\citealtads}[3][][]{\href{http://adsabs.harvard.edu/abs/#3}%
    {\def\hyper@linkstart##1##2{}%
     \let\hyper@linkend\@empty\citealt[#1][#2]{#3}}}
  \newcommandtwoopt{\citeyearads}[3][][]%
    {\href{http://adsabs.harvard.edu/abs/#3}
    {\def\hyper@linkstart##1##2{}%
     \let\hyper@linkend\@empty\citeyear[#1][#2]{#3}}}
\makeatother

\usepackage{subcaption}

\usepackage{mathtools, amsmath}
\usepackage{nccmath}
\usepackage{float}
\usepackage{graphicx}
\usepackage{hyperref}
\usepackage{placeins}
\usepackage{ulem}
\usepackage{hhline}

\usepackage{svg}
\newcommand{\linkorcid}[1]{\href{https://orcid.org/#1}{\includegraphics[width=8pt]{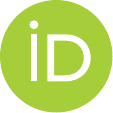}}}

\usepackage{xcolor}
\hypersetup{
    colorlinks,
    linkcolor={red!50!black},
    citecolor={blue!50!black},
    urlcolor={blue!80!black}
}
\usepackage{natbib}

\newcommand{\hmpc }{$h^{-1}$Mpc}
\newcommand{\ihmpc}{$h\, \text{Mpc}^{-1}$}

\newcommand{\sn}{SN~Ia}
\newcommand{\sns}{SNe~Ia}
\newcommand{\kms}{km~s$^{-1}$}

\usepackage[switch]{lineno} 

\renewcommand{\arraystretch}{1.3}

\begin{document}

   \title{Forecast for growth-rate measurement using peculiar velocities from LSST supernovae}

   \subtitle{}

    \author{
        D. Rosselli \inst{\ref{CPPM}}\fnmsep\thanks{Corresponding author \email{rosselli@cppm.in2p3.fr}}\linkorcid{0000-0001-6839-1421}
        \and B. Carreres \inst{\ref{Duke}}\linkorcid{0000-0002-7234-844X} 
        \and 
        C.~Ravoux\inst{\ref{LPC}}\linkorcid{0000-0002-3500-6635}
        \and
        J.~E.~Bautista\inst{\ref{CPPM}}\linkorcid{0000-0002-9885-3989}
        \and D.~Fouchez\inst{\ref{CPPM}}\linkorcid{0000-0002-7496-3796 }
        \and A.G. Kim\inst{\ref{LBNL}
        }
        \and B.~Racine\inst{\ref{CPPM}}\linkorcid{0000-0001-8861-3052}
        \and
        F.~Feinstein\inst{\ref{CPPM}}\linkorcid{0000-0001-5548-3466}
        \and B.~Sánchez\inst{\ref{CPPM}}\linkorcid{0000-0002-8687-0669}
        \and A. Valade\inst{\ref{CPPM}}
        \and The LSST Dark Energy Science Collaboration
          }
         \institute{
                    Aix Marseille Univ, CNRS/IN2P3, CPPM, Marseille, France\label{CPPM}
                    \and Department of Physics, Duke University, Durham, NC 27708, USA\label{Duke}
                    \and Université Clermont-Auvergne, CNRS, LPCA, 63000 Clermont-Ferrand, France\label{LPC}
                    \and Lawrence Berkeley National Laboratory, 1 Cyclotron Road, Berkeley, CA 94720, USA\label{LBNL}
        }


 
  \abstract
  {In this work we investigate the feasibility of measuring the cosmic growth-rate parameter, $f\sigma_8$, using peculiar velocities (PVs) derived from Type Ia supernovae (SNe Ia) in the Vera C. Rubin Observatory’s Legacy Survey of Space and Time (LSST). We produce simulations of different SN types using a realistic LSST observing strategy, incorporating noise, photometric detection from the Difference Image Analysis (DIA) pipeline, and a PV field modeled from the Uchuu Universe Machine simulations. We test three different observational scenarios, ranging from ideal conditions with spectroscopic host galaxy redshifts and spectroscopic supernova typing to realistic photometric supernova typing resulting in contamination with non-Ia SNe. Using a maximum likelihood technique, we show that LSST can measure $f\sigma_8$ with a precision of $10 \%$ in the redshift range $0.02 <z<0.14$ for our most realistic scenario. Using three tomographic bins, LSST will be able to constrain the growth-rate parameter with errors below $18\%$ up to redshift $z=0.14$. We also test the impact of contamination on the maximum likelihood method and we find that for a level of contamination fraction below $\sim 2\%$ we recover unbiased measurements.
  The results of this analysis highlight the potential of the LSST SN sample to complement traditional redshift-space distortion measurements at high redshift, providing a novel avenue for testing general relativity and different dark energy models.}

    \titlerunning{Forecast for growth-rate measurement using peculiar velocities from LSST supernovae}
    \authorrunning{D. Rosselli et al. }
   \maketitle
%

\section{Introduction}

The $\Lambda$-cold dark matter ($\Lambda$CDM) model is widely accepted as the standard cosmological framework that satisfactorily describes the Universe on the largest scales. 
This model assumes that general relativity (GR) is valid at all scales and the existence of cold dark matter (CDM) to explain the growth of structures as well as the rotation curves of galaxies. It also requires dark energy with the dynamics of a cosmological constant, $\Lambda$, to drive the observed accelerated expansion of the Universe.
The nature of dark energy is still unknown, which has led to the development of alternative models of gravity that do not require this component. These alternative models usually predict the same background evolution as the standard cosmological model, while the predictions for the parameters related to the evolution of density perturbations (e.g. the linear growth rate of structures, $f(z)$) are different.

Over the years, GR has undergone rigorous testing on both small and large cosmological scales. However, current measurements are not precise enough to distinguish among the various alternative theories of gravity proposed to explain the accelerated expansion of the Universe and the growth of cosmic structures. Ongoing and future surveys (e.g. the Vera C. Rubin Observatory Legacy Survey of Space and Time, LSST, \citealt{LSST_2009_sciencebook}, Dark Energy Spectroscopic Instrument, DESI, \citealt{desicollaborationDESIExperimentPart2016}, Euclid, \citealt{laureijs2011euclid}, Zwicky Transient Facility, ZTF, \citealt{graham_ztf_2019}) aim to reach the precision that will allow us to discriminate among different models of gravity. The constraints of the expansion rate, $H(z)$, coming from standard candles (type Ia supernovae, \sns\,, e.g. \citealt{descollaboration2024darkenergysurveycosmology}) or standard rulers (Baryon Acoustic Oscillations, BAO, e.g. \citealt{desicollaboration2024desi2024vicosmological}) will reach sub-percent level precision while measurements of $f(z)$ from redshift-space distortions (RSD) will reach percent level precision \citep{desicollaborationDESIExperimentPart2016,amendolaFateLargescaleStructure2018}. This represents a significant improvement over current constraints, which have typical uncertainties of $10\%$ on $f\sigma_8$, as reported in recent analyses (e.g. \citealt{Alam_2021}).

The growth rate of structures is commonly measured through RSD in the clustering of galaxies. This effect is due to the fact that galaxy peculiar velocities (PVs) cause distortions in the observed redshift space with respect to the comoving space. 
The observed redshift, $z_\mathrm{obs}$, of the galaxy is impacted by the PV component along the line-of-sight, $v_p$, through the Doppler shift ($z_p$):

\begin{equation}
    1+z_\mathrm{obs}=(1+z_\mathrm{cos})(1+z_p),
    \label{eq:zcomp}
\end{equation}
where $z_\mathrm{cos}$ is the cosmological redshift, and at the first order $z_p = v_p / c$.
In galaxy surveys the observed redshift is used to infer galaxy distances given the cosmological model, then the shifts with respect to $z_\mathrm{cos}$ due to PVs produce distances which are slightly different from the true comoving ones. The shifts in the inferred distances cause anisotropies in the two-point statistics of the galaxy density field \citep{kaiserClusteringRealSpace1987}. The amplitude of this anisotropy is proportional to $f(z)$ and the amplitude of matter fluctuations, $\sigma_8$ (the standard deviation of the matter density field that has been top-hat smoothed on a scale of $8$ \hmpc). Therefore the measurements of growth rate from RSD usually quote the combination $f(z)\sigma_8(z)$.

It is also possible to measure $f\sigma_8$ by studying the statistical properties of the velocity field itself \citep[see][]{gorsky1989}, instead of its effect on the observed density field statistic. 
PVs can be estimated if both the galaxy redshifts and the absolute distances are measured independently. We can infer the cosmological redshifts from the distances, by assuming a cosmological model, and estimate the PV contribution in the observed redshift. Spectroscopy provides precise redshifts, while galaxy distances can be estimated using relations between galaxy properties and their intrinsic luminosity like the Tully-Fisher relation for spiral galaxies \citep[TF,][]{tullyNewMethodDetermining1977} and the Fundamental Plane for elliptical ones  \citep[FP,][]{djorgovskiFundamentalPropertiesElliptical1987}. 
Another popular distance indicator is \sns\,. Their peak luminosity can be recovered with a $\sim 14\%$ magnitude scatter, yielding a $\sim 7\%$ uncertainty on the distance, compared to $20-30\%$ for TF and FP galaxies, therefore providing more precise PVs.
Until now SN samples have suffered from the limited sky coverage of past surveys or from being a collection of inhomogeneous data from different telescopes (e.g. Pantheon+ sample \citealt{scolnicPantheonAnalysisFull2022}). 
This has limited the use of PVs from \sns\ as a tool to measure the growth rate, and only a few results have been obtained. However, current and future photometric surveys, such as the ZTF and LSST, are expected to provide a sufficiently uniform and large sample of \sns\ that can be used for peculiar velocity studies \citep{howlettCosmologicalForecastsCombined2017,carreresGrowthrateMeasurementTypeIa2023}.

To obtain the constraints on $f\sigma_8$, the statistical properties of a peculiar velocity sample can be measured alone or in combination with an overlapping galaxy sample. The maximum likelihood method (used in this work) is widely used for extracting growth-rate measurements. This method assumes that velocity or density fields are drawn from multivariate Gaussian distributions \citep{johnson6dFGalaxyVelocity2014,adamsJointGrowthRate2020,laiUsingPeculiarVelocity2023,carreresGrowthrateMeasurementTypeIa2023}. Other methodologies use compressed two-point statistics, such as the two-point correlation function, power spectrum, or average pairwise velocities \citep{nusserVelocityDensityCorrelations2017,turnerLocalMeasurementGrowth2022,qin2024redshiftspacemomentumpowerspectrum}.
Furthermore, it is possible to measure $f\sigma_8$ by directly comparing the observed velocities to those reconstructed from a density field \citep{carrickCosmologicalParametersComparison2015,saidJointAnalysis6dFGS2020}. Finally, there are newly developed methodologies such as: field-level inference, and forward modeling approaches to infer the late time velocity and density field (e.g. \citealt{Valade_2022}) or to go back to initial conditions (e.g. \citealt{prideaux-gheeFieldBasedPhysicalInference2022}). We highlight that PVs are particularly powerful probes for the growth rate at low redshift, where the error of $f\sigma_8$ from RSD measurements is dominated by cosmic variance. However, at high redshift, RSD is the most powerful probe, where observed PVs are dominated by measurement errors.
Combining peculiar velocities from SNe Ia, Tully-Fisher and Fundamental plane with galaxy density data from current and future surveys will place the best constraints on the growth rate at low redshift and will allow us to discriminate between GR and different gravity theories \citep{kimComplementarityPeculiarVelocity2020}.

In this work, we study the feasibility of the growth-rate measurement using only SN~Ia data from LSST. We produce realistic simulations of LSST SN~Ia light curves, including selection effects and instrumental noise, and we perform the analysis to measure $f\sigma_8$.
This article is organized as follows. In Sect.~\ref{lsstsim} we describe the simulations of LSST SN light curves. In Sect.~\ref{method} we present the maximum likelihood method. In Sect.~\ref{result} we describe our main results, and in Sect.~\ref{bias_contamination} we investigate the systematic effect of contamination on the growth-rate measurement. Finally, we discuss our results in Sect.~\ref{discussion}, and we state our conclusions in Sect.~\ref{conclusion}.

\section{LSST supernova simulations for peculiar velocities}\label{lsstsim}

LSST is a wide field survey ($\sim 20,000$ deg$^2$) that is planned to start at the end of $2025$ and will go on for 10 years. LSST will be carried out at the Vera C. Rubin Observatory, which is located on the Cerro Pachón in Chile. LSST will observe the sky using an $8.4$ m primary mirror with a nearly $10$~deg$^2$ field-of-view in six passbands (\textit{ugrizy}).
The main programs of the survey will be the Wide-Fast-Deep (WFD) survey, which will cover about $18,000$~deg$^2$, the Galactic plane, polar region surveys, and the Deep Drilling Fields (DDFs). The latter are $5$ smaller sky patches that will be observed with a higher cadence and more depth with respect to the WFD. The WFD and DDF surveys will use a rolling cadence \citep[see][]{LSST_obs_strategy_whitepaper}. Thanks to this observation strategy, LSST will deliver $\sim 10^7$ transient detections over 10 years, which will be made public to the community.


\subsection{Galaxy mocks}\label{uchuu_mocks}

To simulate \sn\ host galaxies that follow a cosmological PV field, we use the Uchuu Universe Machine simulation \citep{ishiyamaUchuuSimulationsData2021,aungUchuuuniverseMachineData2023}. The Uchuu simulation consists of a box with a volume of (2 Gpc~$h^{-1}$)$^3$ available as a snapshot at several redshifts. The simulation contains galaxies along with their properties, obtained from the halos which are found in the initial dark halo simulation, using the \texttt{UniverseMachine} algorithm \citep{Behroozi_2019_Univmachine}. The Uchuu simulation uses the cosmological results from \citet{planckcollaborationPlanck2015Results2016} as fiducial cosmological parameters. In this work, we use the snapshot box at a redshift $z=0$. We subdivide the main box into 8 sub-boxes of $\sim 1,000$~\hmpc\ side-length corresponding to a maximum redshift of $z \sim 0.17$ for a central observer, since that is the relevant redshift range for PV measurements.
Since the 8 mocks in redshift space are derived from a single snapshot at $z=0$, there is no redshift evolution of the cosmological parameters.

    

\subsubsection{Galaxy properties}\label{uchuu_gal_prop}

The Uchuu-UniverseMachine galaxy catalog includes a variety of baryonic properties for all galaxies down to about $5 \times 10^8$ M$_{\odot}$. The galaxy–halo relationship is forward-modeled to match observational data across cosmic times. The star formation rate is parametrized as a function of halo mass, halo assembly history, and redshift. The stellar mass of the galaxy hosted by the halo is computed by integrating the star formation rates over the merger history of the halo, accounting for mass lost during stellar evolution.
We complement the information inside the Uchuu-UniverseMachine catalog by adding the galaxy magnitude in various filters (including the LSST ones) and the galaxy Sérsic profile parameters. The extra information is necessary to compute the noise from the host galaxy surface brightness in the SN fluxes.

To add these properties we use the \texttt{OpenUniverse} LSST-Roman simulations \citep{openuniverse_openuniverse2024_2025}. The \texttt{OpenUniverse} is a simulated synthetic imaging survey in a $20$ deg$^2$ overlap between the Nancy Grace Roman Space Telescope High Latitude Wide Area Survey (HLWAS) and five years of observations of LSST. The LSST-Roman synthetic images are created with reference to the existing $300$ deg$^2$ of LSST simulated imaging produced as part of the Dark Energy Science Collaboration (DESC) Data Challenge 2 (DC2) \citep{desc_dec2_2021}. DESC DC2 images are based on the cosmoDC2 synthetic galaxy catalog \citep{Korytov_2019_cosmoDC2}. The LSST-Roman simulation contains a non-magnitude limited sample of galaxies with their observed magnitudes for LSST filters (\textit{ugrizy}), SDSS filters (\textit{ugriz}), and Roman photometric bands (\textit{R062, Z087, Y106, J129, W146, H158, F184, K213}). This catalog also includes the star formation rate and the stellar mass associated with each galaxy. Therefore, we perform a two-dimensional interpolation on the LSST-Roman catalog over the masses and star formation rate parameters to assign the realistic galaxy magnitudes and and the galaxy Sérsic profile parameters to Uchuu galaxies. 

\subsection{LSST observing conditions}\label{OPSIM}

To simulate the LSST SN light curves we make use of the baseline version 3.3 observing strategy simulation from the Operation Simulator\footnote{\url{https://www.lsst.org/scientists/simulations/opsim}}(\texttt{OpSim}, \citealt{delgado_2014_opsim}). \texttt{OpSim} simulates the LSST observed fields over the 10-year survey. It takes into account observing conditions with a time-dependent model of seeing, cloudiness, moon brightness, and a detailed model of the telescope and the dome. These models are based on data taken over various years at the Rubin Observatory site. \texttt{OpSim} creates a catalog with the sky coordinates of the telescope pointings, the time of every observation, the filters, and the information that describes the observing conditions such as air mass, point spread function (PSF), sky brightness, and $5\sigma$ magnitude depth.

\begin{figure}
    \centering
    \includegraphics[width=\hsize]{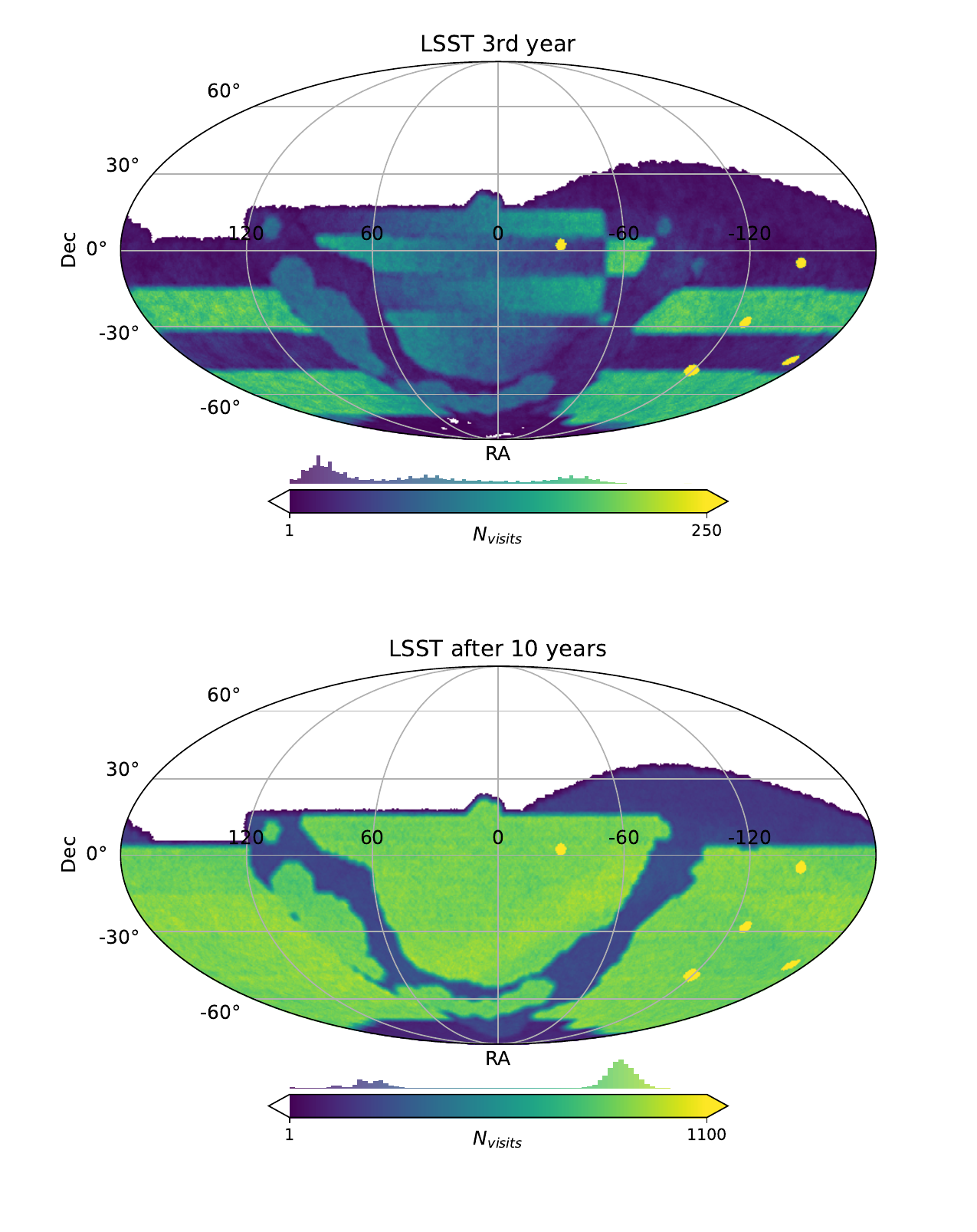}
    \caption{LSST survey footprint showing the number of visits ($N_{\rm visits}$) in each sky location. The top panel shows the visits during the third year when the rolling cadence is active, while the bottom one shows the survey as expected after completion (10 years of observations).}
    \label{fig:LSST_obs_sky}
\end{figure}

Figure~\ref{fig:LSST_obs_sky} shows the LSST survey footprint and the number of visits in each sky location from the \texttt{OpSim} baseline version 3.3. It shows the observation during the third year of the survey (top panel), when the rolling cadence is active, and the status of the survey after its entire duration (bottom panel). After 10 years LSST is expected to be homogeneous in the WFD area ($\langle N_{\rm visits} \sim 900 \rangle$); the DDFs area (yellow spots) will be observed more ($\langle N_{\rm visits} \sim 23,000 \rangle$), while the Galactic plane region will be observed less. During the rolling cadence, there will be active regions that will be observed more (higher cadence) compared to the background areas.

\subsection{Supernova simulations}\label{snesim}

We generate synthetic SN light curves using the \texttt{snsim}\footnote{\url{https://github.com/bastiencarreres/snsim}} python library, as previously done in \citet{carreresGrowthrateMeasurementTypeIa2023} and \citet{carreres2024ztfdr2} This code allows us to simulate realistic LSST SN light curves and parameters on top of the mocks described in Sect.~\ref{uchuu_mocks}. The SN host galaxies are randomly selected inside the LSST footprint from the Uchuu mocks. 
The SN light curves are generated starting from a spectral energy distribution (SED) model including astrophysical effects: PVs effect on the redshifts, beaming due to the effect of PVs \citep{davisEffectPeculiarVelocities2011}, and Milky Way extinction.\footnote{ CCM89 model \citealt{cardelliRelationshipInfraredOptical1989} dust law with $RV = 3.1$ and EBV from \citealt{schlegelMapsDustInfrared1998}.} Hence, the SN observed redshifts and magnitudes are affected by correlated PVs due to the large-scale structure of each galaxy mock. In this analysis, we neglect any type of correlation between the SN rates and the host-galaxy properties (e.g. mass, star formation rate, star formation history). The effect of these correlations on the $f\sigma_8$ measurement will be studied in future works.

We generate the true light-curve fluxes for each SN by integrating the SED models over the LSST filters.\footnote{For this operation we use the \texttt{SNCOSMO} package \citealt{barbarySNCosmo2023} (\url{https://sncosmo.readthedocs.io/}).}
The true fluxes are converted into observed fluxes and relative uncertainties using the observing conditions given by the OpSim simulation, described in Sect.~\ref{OPSIM}. The flux uncertainties are estimated as the quadratic sum of the sky noise, the Poisson noise from the source, the noise from the host-galaxy surface brightness and the error on the zero point (ZP) estimation. We set $\sigma_{\mathrm{ZP}} = 0.005$ mag as a calibration uncertainty; this uncertainty will be refined in further works, together with studies on the impact of calibration uniformity on the $f\sigma_8$ measurement. 

In the following subsections, we describe the details and assumptions used to simulate different types of SNe.

\subsubsection{"Normal" SNe Ia}

Our simulation is primarily composed of "normal" SNe Ia, or just SNe Ia, the supernovae that are standardizable candles and are used in cosmological analyses.
To simulate the normal \sns\ we utilize the \texttt{SALT3} SED model \citep{salt3_kenworthy_2021}. 
The \sns\ are randomly assigned to the host galaxies in the Uchuu mocks following the redshift distribution given by the volumetric rate presented in \citet{Frohmaier_PTF_2019_sniarate}, which is rescaled for the value of $H_0$ in the Uchuu simulation.
In our simulations we model the stretch distribution, $x_1$, following the redshift-dependent two-Gaussian mixture from \citet{nicolasRedshiftEvolutionUnderlying2021} and the color distribution, $c$, using the asymmetric model described in \citet{scolnicMeasuringTypeIa2016}. 
The SN~Ia intrinsic scattering is drawn from a normal distribution with a fixed dispersion equal to $\sigma_M=0.12$ mag. The analysis of the effect of different intrinsic scatter models, such as the color-dependent one described in \citet{brout_scolnic_2020_colorscattering}, is presented in \citet{carreres2025typeiasupernovagrowthrate}.
Finally, for each \sn\ we compute the apparent magnitude using the Tripp formula \citep{trippTwoparameterLuminosityCorrection1998}:

\begin{equation}\label{tripp}
    \mu_i = m_{B,i} + \alpha x_{1,i} - \beta c_{i} - M_0,
\end{equation}
where $\mu_i$ is the distance modulus of each \sn. $\alpha$ and $\beta$ are the coefficients that relate the stretch and the color to the magnitude for the whole SN Ia population. $M_0$ is the SN Ia intrinsic brightness, which is rescaled by the $H_0$ value in the Uchuu simulation. The values of the Tripp input parameters in the simulation are: $\alpha=0.14$, $\beta=2.9$, and $M_0=-19.12$ mag.

    

\subsubsection{Peculiar SNe Ia}

We also simulate two types of peculiar \sns\ that are contaminants in \sn\ samples: SN1991bg-like SNe (\citealt{filipenko_sni91g_1992}, hereafter SNe 91bg-like) and SN2002cx-like supernovae (\citealt{Li_2003_sniax,foley_2013_sniax}, hereafter SNe Iax).

SNe 91bg-like are sub-luminous compared to normal SNe Ia, they are characterized by smaller $x_1$ (fast-declining), and are redder at peak with respect to normal SNe Ia. In our simulations, we used the corrected SED library of 35 91bg-like events from the Photometric LSST Astronomical Time Series Classification Challenge\footnote{\url{https://zenodo.org/records/6672739}} \citep[PLAsTiCC;][]{Kessler_2019_plasticc}. As in \citet {Kessler_2019_plasticc}, we assume that the SN 91bg-like volumetric rate is $12 \% $ of the SN~Ia rate.

SNe Iax generally rise and decline faster than normal SNe Ia. Again, we use the models presented in PLAsTiCC, which are based on SN 2005hk \citep{Phillips_2007}. As in \citet{vincenzi_2021_descontamination}, to reproduce the color diversity of SNe Iax we apply a range of host extinctions in the simulations following the host-extinction distribution described in \citet{rodney_2014}. Following \citet{Kessler_2019_plasticc}, we assume the volumetric rate of SNe Iax at $z=0$ to be $24 \%$ of the normal \sn\ rate and the redshift evolution of the SN Iax rate is chosen to follow the cosmic star formation rate from \citet{madau_dickinson_2014}.

\subsubsection{Core-collapse SNe}

Our simulation of core-collapse SNe (hereafter SNe CC) is based on the library of $67$ SED templates presented in \citet{Vincenzi_2019_templates}. This library is created by combining spectroscopy and multi-band photometry of SNe CC across $6$ different sub-classes (SNIIP/L, SNIIb, SNIIn, SNIb, SNIc, and SNIc-BL). 
For the relative rate of each SN CC sub-type, we use the measurements described in \citet{Shivvers_2017}. To simulate the redshift distribution we assume that the SN CC rate follows the cosmic star formation history
from \citet{madau_dickinson_2014} normalized by the local SN rate measured by \citet{Frohmaier_2021}.
We normalize the brightness of the SN CC templates using the Gaussian fit shown in Table 5 of \citet{vincenzi_2021_descontamination}. The values of the luminosity function for each SN CC sub-type are computed in \citet{Vincenzi_2019_templates} using the measurement performed by \citet{LI_2011} and the revised rates from \citet{Shivvers_2017}. In our simulation, we use the set of templates from \citet{Vincenzi_2019_templates} that have not been corrected for host-galaxy dust extinction because the luminosity functions are also measured from SNe that were not corrected for host-galaxy dust extinction. 

\subsubsection{\texttt{snsim} output}


\texttt{snsim} outputs the observed fluxes and simulated parameters of each SN. 
In simulating LSST, we apply a cut on the sampling of the light curves. We simulate the SNe with at least $10$ observations within $-50 < t_0 < 150$ days in any band. This selection is applied to ensure a minimum sampling of the light curves, speed up the simulations and reduce the data storage. We perform tests changing this initial cut and conclude that this selection does not affect the results of this analysis. Moreover, by computing the selection function of the simulation output with respect to the parent sample given by the assumed rate, we verify that the initial simulated sample is complete in redshift. 
The average number of SNe in our LSST simulations is $\langle N \rangle = 318,579 $.
We highlight that all the results that we show in this work strongly depend on the assumptions made for the simulation, especially on the rates for the different SN types. To assess the real impact of contamination and selections on the growth rate, it will be necessary to calibrate the simulations on the first year of LSST data. This means creating simulations that reproduce the parameter distributions of the real data, as done in \citet{vincenzi_2021_descontamination}.


\subsection{Light-curve selection}\label{selection_DIA}

After simulating the SN light curves, as described in the previous section, we apply a series of selections to construct the expected sample that will be detected by LSST:

\begin{enumerate}
    \item We remove all the saturated observations using the nominal LSST saturation limit\footnote{$u, g, r, i, z, y = 14.7, 15.7, 15.8, 15.8, 15.3, 13.9$ mag values of LSST nominal saturation limit at 0.7” seeing from \url{https://www.lsst.org/sites/default/files/docs/sciencebook/SB_3.pdf}} (we do not account for changes in the saturation limit due to different observing conditions, e.g. different seeing conditions).
    \item We apply the detection probability expected from the Difference Image Analysis (DIA) LSST-DESC pipeline. We use the detection efficiency as a function of the signal-to-noise ratio (SNR), as shown in Figure~9 of \citet{Brunooooo_2022}. We keep in our sample all the light curves that have at least one detected observation in any band.
    \item To determine which SNe will have a host galaxy with measured spectroscopic redshift, we apply host spectroscopic redshift efficiencies from the Dark Energy Survey Instrument (DESI) and the 4-metre Multi-Object Spectroscopic Telescope (4MOST). In Sect.~\ref{host_spectra_selection} we describe how those efficiencies have been computed. Naturally, we select only the SNe that are observed by LSST and lie inside the DESI and 4MOST footprints. 
    \item We coadd the observations per night. Some light curves have two or more separated observations taken in the same night through the same filter, so we merge those observations into a single one and recompute the SNR. Finally, we keep in the sample the light curves that have SNR~$>4$ in at least three distinct bands.
\end{enumerate}

\subsubsection{Host spectroscopic redshift selection}\label{host_spectra_selection}

We determine the efficiencies of the DESI Bright Galaxy survey Sample (BGS), 4MOST Cosmology Redshift Survey bright galaxies (CRS-BG) and 4MOST Hemisphere Survey of the Nearby Universe (4HS) surveys to get the spectroscopic redshifts of SN host galaxies. 
We define the SN/host redshift efficiency, $R_{\mathrm{host}}$ as the ability of a spectroscopic survey to measure the redshift of an LSST SN host. For each SN type, the efficiency can be mathematically characterized by the number of LSST SN hosts whose redshifts are measured normalized by the total number of SNe. We assume that $R_{\mathrm{host}}$ can be decomposed as:

\begin{equation}\label{eq:Rhost}
   R_{\mathrm{host}}(z) =  R_{\mathrm{spectro}}(z) \times R_{\mathrm{SN}}(z) ,
\end{equation}
where $R_{\mathrm{spectro}}$ is the survey redshift efficiency and $R_{\mathrm{SN}}$ is the proportion of SNe located in the targeted hosts with respect to the total number of SNe.

To compute $R_{\mathrm{host}}$ we use the DESC DC2 simulations \citep{desc_dec2_2021} and, in particular, the associated LSST-Roman \texttt{OpenUniverse} catalog \citep{openuniverse_openuniverse2024_2025}, described in Sect.~\ref{uchuu_gal_prop}.

For each survey, we use the corresponding magnitude and color cuts that define the target galaxy sample. The target galaxies are similar for the three considered surveys, since at low redshifts all the surveys target Bright Galaxies (BGs). We apply the target selection cuts on the galaxies inside cosmoDC2 simulation. By normalizing the galaxy density as a function of redshift inside cosmoDC2, $n_{DC2}(z)$, by the expected $n(z)$ of each survey, we estimate the corresponding survey redshift efficiency, $R_{spectro}$ as a function of redshift. The DESI $n(z)$ is obtained from the public information of the target selection in \citet{hahn_desi_2023}. Further information about target selection for the 4MOST surveys is obtained from \citet{4most_4hs} and \citet{4most_crs_bg} and complemented with the expected $n(z)$.\footnote{Private communication from E. N. Taylor and J. Kneib, the principal investigators of each survey}
We use the galaxy properties in the simulation, such as stellar mass and star formation rate, to simulate the SNe and assign them to the most probable host. We apply the galaxy target selection cuts of each survey to the SNe simulated in the cosmoDC2 simulation and we normalize by the total number of SNe. This provides an estimation of the proportion of SNe located in the targeted host, $R_{\rm SN}$. The details on the SN simulation inside cosmoDC2 and on the target selection of each survey are described in Appendix \ref{app:r_host_details}.

Finally, we compute $R_{\rm host}$ following Eq.\eqref{eq:Rhost} for each survey. We compute $R_{\rm host}$ only as a function of redshift, assuming that it does not vary inside the survey footprints.\footnote{The code used to compute the SN/host efficiencies for this work is publicly available at \url{https://github.com/corentinravoux/desidescsn}}

\begin{figure}
    \centering
    \includegraphics[width=\hsize]{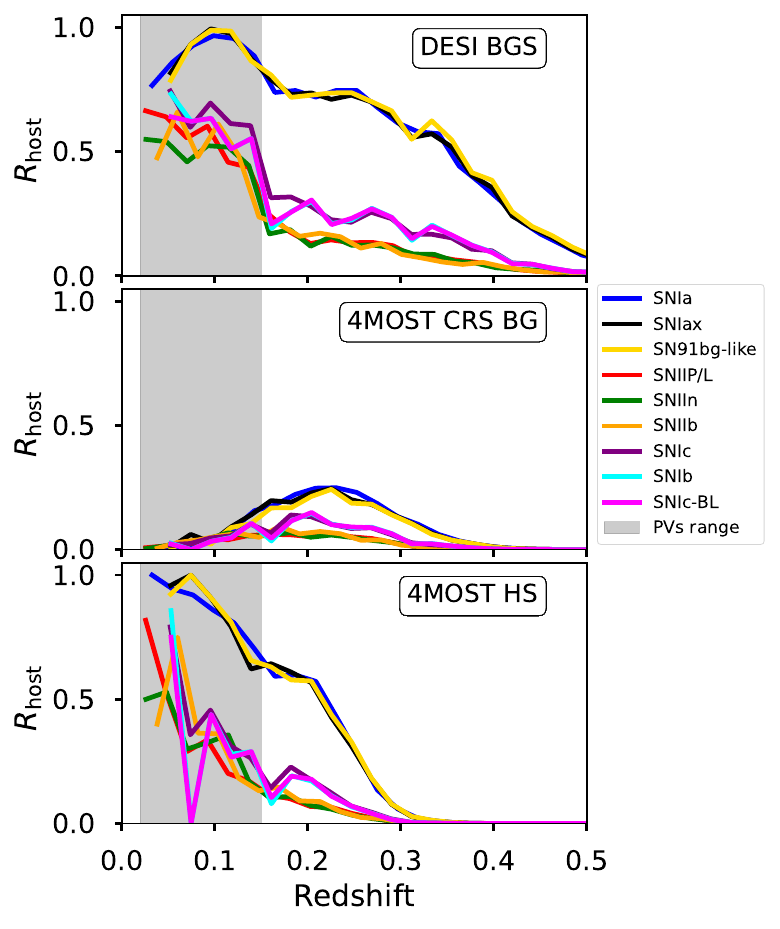}
    \caption{SN/host redshift efficiencies, $R_{\mathrm{host}}$, as a function of redshift for each type of SN in the simulation. The top panel shows the efficiencies for DESI BGS survey, the central panel shows the ones for 4MOST CRS BG survey and the bottom panel shows the efficiencies for 4MOST 4HS surveys. The gray shaded area shows the redshift range of interest for LSST PV measurement.}
    \label{fig:host_z_eff}
\end{figure}

The SN/host redshift efficiencies of DESI BGS, 4MOST CRS-BG and 4HS for all the SN types in our simulations are given in Fig. \ref{fig:host_z_eff}, which does not show the number of hosts observed but rather the proportion. A low proportion can represent a large number of host-galaxy redshifts, especially at high redshifts where the volume is larger. The efficiencies are different for each SN type since different types of SNe tend to occur in galaxies with distinct properties. 
We note that both DESI BGS and 4MOST HS have a high efficiency (almost one) for SNe Ia at $z<0.1$. The high efficiency means that for $z<0.1$ we will be able to get the spectroscopic redshift for almost all the SN Ia host galaxies, provided they are in the footprint of the surveys. 

\subsection{Photometric classification}\label{photometric_classification}

After applying the selections described above, the next step of the analysis is the SN classification. LSST will discover an unprecedented number of supernovae, exceeding the resources available to acquire SNe spectra. The 4MOST Time-Domain Extragalactic Survey (TiDES, \citealt{4most_tides}) will provide spectroscopic follow-up only for $\sim$~$30\,000$ SNe at all redshifts \citep{frohmaier2025tides4mosttimedomain}. Hence, a large part of the LSST SNe will not have a spectroscopic classification. LSST will rely on photometric classifiers: machine learning models trained on light curves to infer the SN types.

In this work, we use the SuperNNova (SNN; \citealt{Moller_2019_SNN}) framework to perform photometric classification of our simulated SN datasets and measure for each SN event its probability of being a SN Ia ($P_{Ia}$). We chose SNN since the code is publicly available\footnote{\url{https://github.com/supernnova/SuperNNova}} and has been used for the Dark Energy Survey (DES) year 5 cosmological analysis \citep{vincenzi2024darkenergysurveysupernova,descollaboration2024darkenergysurveycosmology}. 
To train SNN we run a new simulation of the LSST 10-year sample using the same inputs described in Sect.~\ref{snesim}. We apply all the selections described in Sect.~\ref{selection_DIA} to have a training sample that is representative of the SNe we want to classify. Since SNN needs a large training sample, simulations have typically been used to train the model \citep[see][]{vincenzi2024darkenergysurveysupernova}.
Because the SALT3 model we use to simulate the SNe Ia is defined in the phase range $-20 < t_0 < 50$, we restrict the light curves for each SN type in our training sample. We define $t_0$ as the time of the observation with maximum flux in any band within all the observations with SNR$ >4$. 
This cut is implemented because the classifier may otherwise learn that any observation outside the range $-20 < t_0 < 50$ indicates the object is not a SN Ia, which is not realistic for real data.
Additionally, we keep in the training sample only the light curves with at least $10$ observations in that phase range, to ensure well sampled light curves.
The final training sample is composed of about $50,000$ SNe Ia and about $18,000$ contaminants. In the case of the binary classification task SNN maximizes the accuracy, so the classifier needs a balanced training sample \citep{Moller_2019_SNN}. SNN automatically downsamples the class with the highest number of objects in the training sample, so at the end, we train the classifier with about $36,000$ objects (half SNe Ia and the other half for all the other types). 
To train the classifier we use the same hyperparameters as \citet{Moller_2019_SNN} and \citet{vincenzi_2021_descontamination} and we add the spectroscopic redshift information. Additionally, we normalize the input fluxes using the "cosmo" flag, where each SN multi-band light curve is normalized independently and the normalization factor is the SN maximum flux in any filter.
In Sect.~\ref{contaminated_sample} we show the classification metrics and results obtained using SNN on our 8 LSST realizations.

\subsection{Light-curve fit and quality cuts}\label{SNefit}

The fit of each SN light curve is performed using the same framework as was used to generate the SNe Ia (\texttt{SALT3}).
For each light curve, we fit the stretch, $x_1$, color $c$, peak apparent magnitude, $m_B$, and time of peak brightness, $t_0$. We assume that the error on the observed redshifts is negligible. We perform the light-curve fit on all observations within $-15 < t_0 < 45$. The covariance matrix $C_{\mathrm{SALT},i}$ of the SALT fit is defined as:
\begin{equation}\label{salt_covariance}
    C_{\mathrm{SALT},i} =
\begin{pmatrix}
\sigma^2_{x_1,i} & \mathrm{COV}_{x_1-c,i} & \mathrm{COV}_{x_1-m_B,i}\\
\mathrm{COV}_{x_1-c,i} & \sigma^2_{c,i} & \mathrm{COV}_{m_B-c,i}\\
\mathrm{COV}_{m_B-x_1,i} & \mathrm{COV}_{m_B-c,i} & \sigma^2_{m_B,i}
\end{pmatrix},
\end{equation}
which is used to fit the standardization parameters and recover the PVs, see Sect.~\ref{PVestimation}.

After the light-curve fits, we apply quality cuts to ensure the robustness of the fit results.\footnote{The quality cuts applied in this work are the ones which are commonly used for cosmological analyses.} Considering the resulting $\chi^2$ and the degrees of freedom of each light-curve fit, we select only the best-fit models describing the data with a probability larger than $95\%$.
To ensure a robust estimate of $m_B$, we select light curves that contain at least three observations within $-10<t_0<10$ days. We discard any SN with best-fit stretch $|x_1| > 3$ or color $|c| > 0.3$. Finally, we exclude objects for which the uncertainty on $t_0$ or $x_1$ is larger than $1$, or the uncertainty on $c$ is larger than $0.05$.

\subsection{The final samples}\label{final_samples}

In the following sections, we describe the final SN samples that we recover after applying the selections we have described in the previous sections. We create different scenarios for the LSST SN survey to perform the $f\sigma_8$ measurement and deliver the forecasts under different conditions. In the next sections we present those scenarios, the assumptions made to construct each of them, and the characteristics of each sample. The samples are presented from the most optimistic to the most realistic one.

Our simulations do not take into account the time needed to generate the galaxy templates for the DIA pipeline. The template generation strategy affects the light-curve quality and the detection efficiency during the first months of the survey \citep{Street_2020_lsst_templates}. Different template generation strategies can slightly change the number of SNe Ia reported in this work.

\subsubsection{Full sample}\label{SNIa_pure_sample_full}

For the most idealized scenario, we create a pure sample of SNe Ia. On this sample we perform the selections described in Sect.~\ref{selection_DIA}, except for the host spectroscopic redshift efficiency (Sect.~\ref{host_spectra_selection}). We perform the SALT fit on all the light curves that have passed our first selection and we apply the cuts on the fit results, as described in Sect.~\ref{SNefit}. In this scenario we assume that LSST will have unlimited external spectroscopic resources to acquire SNe Ia and their host spectra all over the footprint. We create this ideal scenario to understand the completeness of the survey given the ability to detect the SNe Ia and to show the full potential of LSST. Hereafter, we call this scenario the Full sample.

\begin{figure}
    \begin{subfigure}[b]{\hsize}
        \centering
        \includegraphics[width=\hsize]{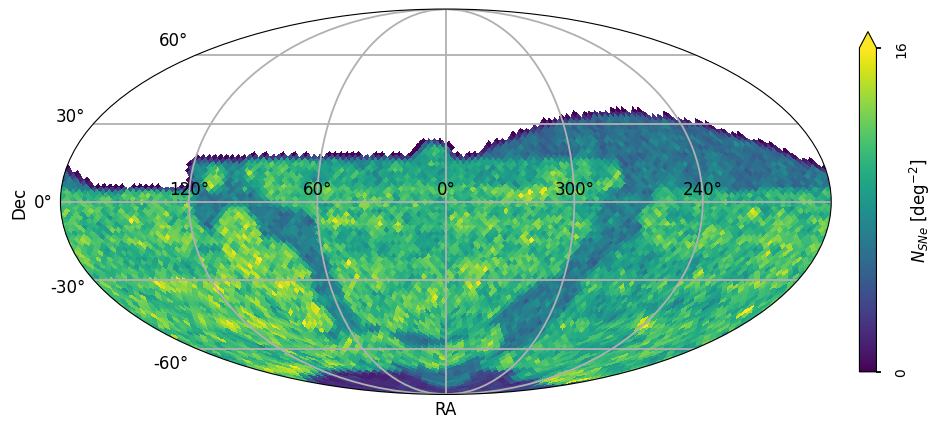} 
        \caption{Density Initial simulation.}
        \label{fig:density_full_sim}
    \end{subfigure}


    \begin{subfigure}[b]{\hsize}
        \centering
        \includegraphics[width=\textwidth]{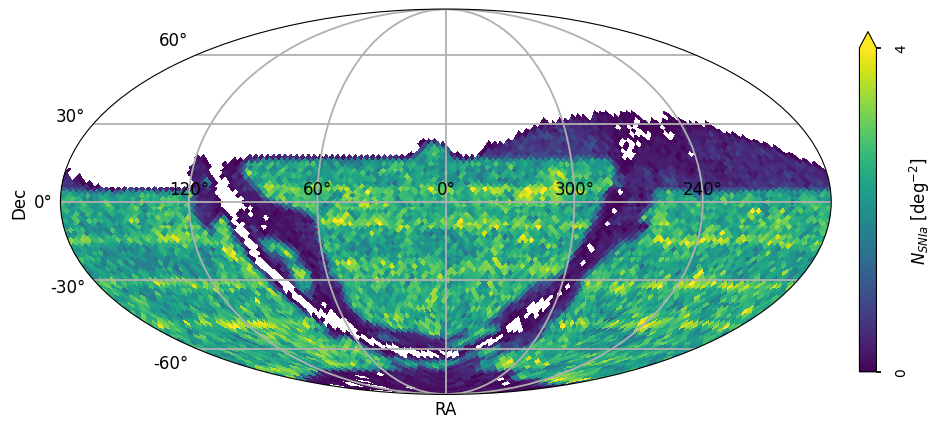} 
        \caption{Density Full sample.}
        \label{fig:density_full_sample}
    \end{subfigure}
    \hfill
    \begin{subfigure}[b]{\hsize}
        \centering
        \includegraphics[width=\textwidth]{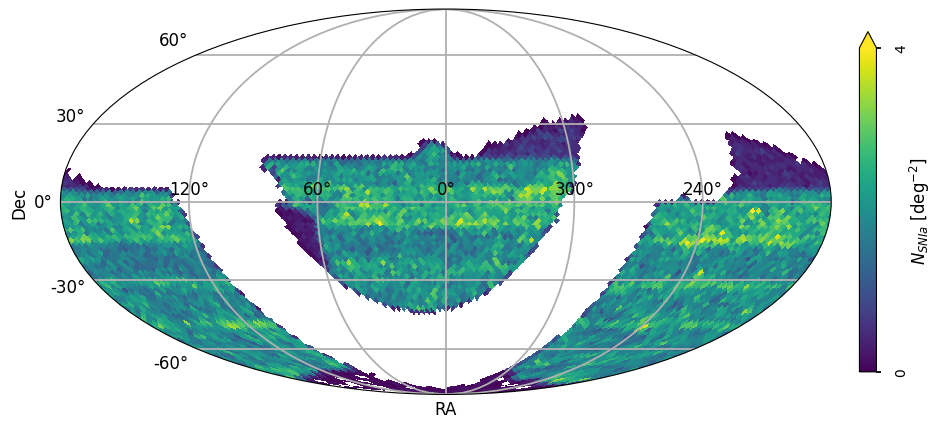} 
        \caption{Density Spec-z sample.}
        \label{fig:density_specz_sample}
    \end{subfigure}



    \caption{Density maps of the different SN Ia samples across the LSST footprint. SN density in the initial simulated sample (a), density of the Full sample (b), and density of the Spec-z sample (c). All the maps represent the average over the 8 LSST realizations.} 
    \label{fig:all_samples+on_sky}
\end{figure}

After all the selections, the average sample over $8$ LSST realizations contains $\langle N \rangle = 52,326$ SNe Ia up to $z=0.16$. We analyze SNe up to $z=0.16$ because the Uchuu mocks have a maximum redshift of $z\sim 0.17$ (Sect.~\ref{uchuu_mocks}), so we cut the SN samples before the mock limits to avoid possible artifacts in the PV field at the mock edges. We remind the reader that the maximum redshift for PV studies is $z\sim 0.2$ because for higher redshifts the PV effect on the redshift is too small to be properly measured \citep{Strauss_1995}.
Figure~\ref{fig:density_full_sample} shows the number density of \sns\ in the Full sample across the sky. Figure~\ref{fig:density_full_sample} shows that the SN~Ia number is uniform inside the WFD area, as expected, while we lose \sns\ around the galactic plane due to the lower light-curve sampling and dust extinction.

\begin{figure}
    \centering
    \includegraphics[width=\hsize]{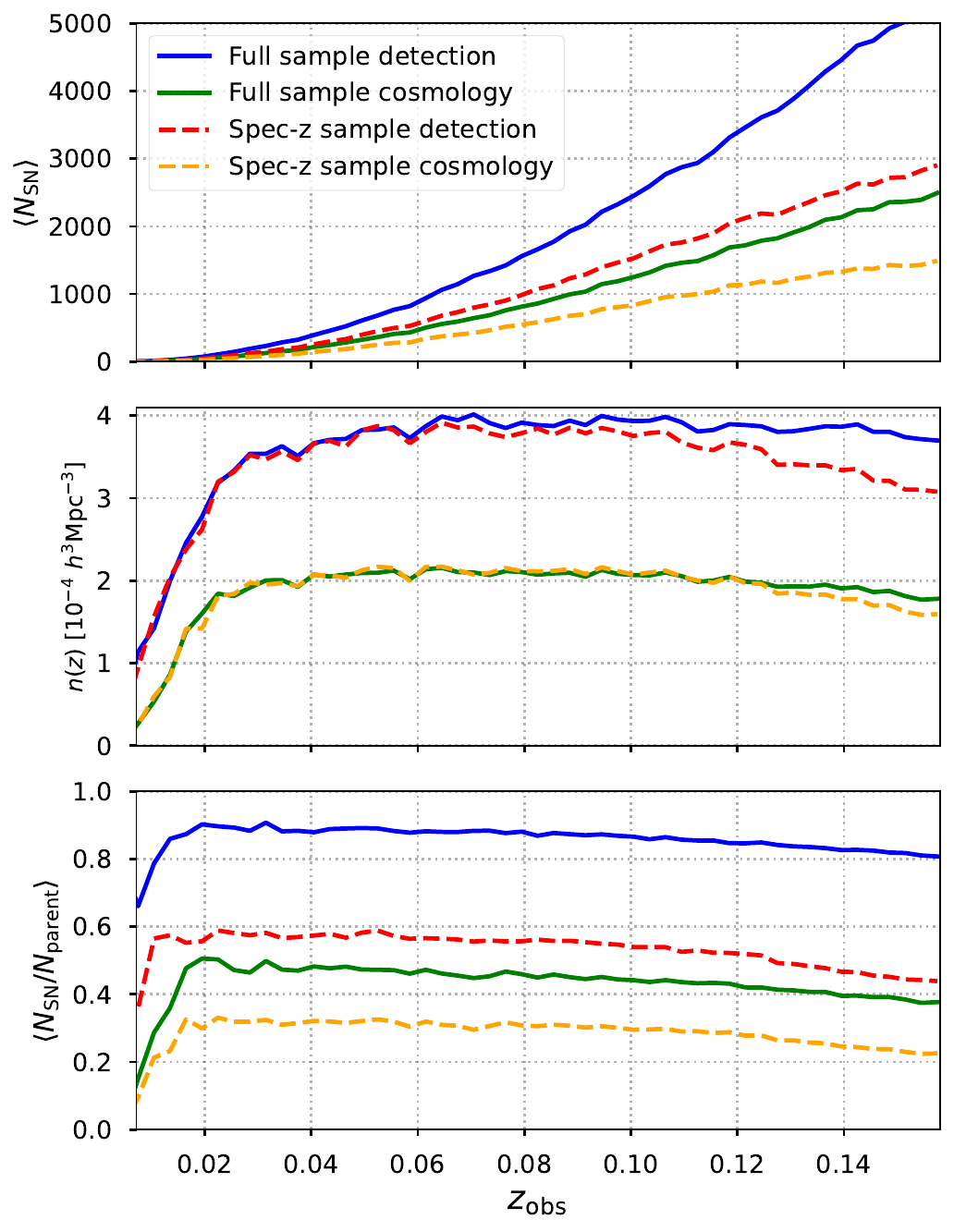}
    \caption{ Redshift distribution (top panel), comoving density (middle panel), and redshift completeness (bottom panel) for the Full and Spec-z samples. The Full sample detection and Spec-z sample detection show the results for the samples before running the SALT fit. The Full sample cosmology and Spec-z sample cosmology present the samples after the quality cuts on the SALT fit results. All the results are the average over the 8 LSST realizations. The completeness distributions are computed with respect to the initial simulated sample (considering only the SNe Ia). }
    \label{fig:red_distr_all_samples}
\end{figure}

Figure~\ref{fig:red_distr_all_samples} shows the SN~Ia average redshift distribution (top panel), comoving density (middle panel), and redshift completeness for the Full sample. The density of the Full sample, both after detection and after SALT fit, increases at low redshift up until $z \sim 0.025$. This means that at extremely low redshift we lose SNe~Ia due to selection effects caused mainly by the saturation of the camera. There are about $100$ SNe Ia with at least one saturated observation for $z < 0.025$. The effect of saturation is confirmed in the redshift completeness plot where we clearly see selections at low redshift. Naturally, if LSST discovers those SNe at early time, before they reach the maximum and saturate, they can be followed up by other surveys such as La Silla Schmidt Southern Survey \citep[LS4,][]{miller2025lasillaschmidtsouthern}. The redshift completeness plot shows that we lose SNe Ia for redshifts higher than $\sim 0.08$. At higher redshifts we lose the dimmer SNe Ia.
In Sect.~\ref{result} we discuss how the selection at higher redshift affects the growth-rate measurement.

\subsubsection{Spec-z sample}\label{SNIa_pure_sample_spectro_selected}

For a more realistic scenario we create a pure sample of SNe Ia, but we apply all the selections of Sect.~\ref{selection_DIA}, including the host spectroscopic redshift. We perform the SALT fit on all the light curves that have passed our first round of selections and we apply the same cuts on the fit results as before. In this scenario we assume that LSST will make use of the observations of DESI and 4MOST to get the spectroscopic redshifts of the \sn\ host galaxies. We also assume that LSST will have unlimited external resources to acquire SN spectra within the DESI and 4MOST footprints.
This scenario is created to deliver a realistic forecast for the growth rate from LSST \sns\ without worrying about possible biases due to contamination. In this case we want to show the realistic statistical power of LSST and to understand the synergies with other facilities. Hereafter, we call this scenario the Spec-z sample. We highlight that in an analysis with real data, a fraction of the \sns\ will be associated with the wrong galaxy, so those \sns\ will be attributed a wrong redshift. In the Dark Energy Survey 5-Year (DES-SN5YR) photometric sample analysis the percentage of host mismatch was found to be about $1.7 \%$, see \citep{HelenQu_DES5yr_hostmismatch}. The mismatch causes a small systematic error in the final cosmology measurement because of the wrong redshifts assigned to the SNe. In this analysis we do not consider any \sn\ host mismatch and we leave the study of the impact of this mismatch on the $f\sigma_8$ measurement to future works.

After all the selections, the average Spec-z sample over $8$ LSST realizations contains $\langle N \rangle = 33\,682$ \sns\ up to $z=0.16$. Figure~\ref{fig:density_specz_sample} shows the number density of SNe Ia across the sky. Figure~\ref{fig:density_specz_sample} shows the \sns\ that lie inside the intersection between the LSST footprint and DESI + 4MOST footprints, thus the region around the galactic plane is completely empty.

Figure~\ref{fig:red_distr_all_samples} shows the  average SN~Ia redshift distribution (top panel), comoving density (middle panel), and redshift completeness for the Spec-z sample. Like the Full sample the density increases at low redshift up until $z \sim 0.025$ due the selection caused by saturation at extreme low redshift. The comoving density of the Spec-z sample is similar to that of the Full sample one up to $z \sim 0.12$. This means that we will have the spectroscopic redshift for almost all the \sn\ host galaxies up to that redshift. In fact, the host-spectrum efficiencies are almost one for DESI BGS and 4MOST HS at low redshift, see Fig. \ref{fig:host_z_eff}. The host-spectroscopic efficiencies start to decrease for $z > 0.12$, so the density of the Spec-z sample is lower than that of the Full sample. The redshift completeness shows that the Spec-z sample also becomes incomplete above $z \sim 0.08$.

In Sect.~\ref{resulsts_specz_sample} we show and discuss our results of the growth-rate measurement for the Spec-z sample.

\subsubsection{Photo-typed sample}\label{contaminated_sample}

For the most realistic scenario, we simulate all of the SN types described in Sect. \ref{snesim}. We apply all the selections described in Sect.~\ref{selection_DIA}, including
the host spectroscopic redshift efficiency, and we classify the SNe using SNN (Sect.~\ref{photometric_classification}). Hereafter, we call this sample the Photo-typed sample.

Within the SNN binary classification framework that separates SNe Ia from the other SN types, we define correctly classified SNe Ia as true positives (TP), while correctly-classified contaminant supernovae are true negatives (TN). Misclassified instances, where SNe Ia are identified as CC or vice versa, are treated as false positives (FP) and false negatives (FN) respectively. The classification is based on the highest-probability prediction, so an object is classified as SN Ia if the probability given by SNN is $P_{Ia} > 0.5$.
To evaluate the effectiveness of our classification method, we focus on two primary performance metrics: contamination fraction and classification efficiency. The contamination represents the fraction of non-SNe Ia among all events classified as SN Ia, and is defined as:
\begin{equation}\label{eq:SNNcontamination}
    \mathrm{Contamination} = \frac{\mathrm{FP}}{\mathrm{FP}+\mathrm{TP}} ,
\end{equation}
so a good classification means low contamination.
The classification efficiency measures the percentage of actual SNe~Ia that were successfully identified and is defined as:
\begin{equation}\label{eq:SNNefficiency}
   \mathrm{Efficiency} = \frac{\mathrm{TP}}{\mathrm{FN}+\mathrm{TP}} ,
\end{equation}
so a good classification means high efficiency.

From SNN classification we initially identify a sample of $\langle N \rangle = 68,084$ SN candidates as likely SNe Ia.
We find a classification efficiency of $92.3 \pm 0.1 \%$, with contamination fraction at $1.15 \pm 0.04 \%$ (these values are the mean over the 8 LSST realizations and the errors are the standard deviations).
By imposing additional quality selection through SALT fit cuts (Sect.~\ref{SNefit}), this sample is refined to $\langle N \rangle = 33,269$ SNe. The resulting contamination fraction for this more stringent sample is $0.021 \pm 0.007 \% $. The small contamination demonstrates the strength of SNN in performing photometric classification for a low-z SN sample.
The contamination fraction is lower than has been seen before. The low contamination can be a results of the fact that we are analyzing only low-redshift SNe, but a proper investigation of this is beyond the scope of this work and we leave it for future studies.

For this sample we do not show the number density across the sky and the SNe redshift distribution because they are not significantly different from the Spec-z sample and all comments given in the previous section are still valid.  

In Appendix \ref{app:SNNvsparsnip} we describe the performance comparison between SNN and the Parsnip classifier \citep{Boone_2021_parsnip}. There we show that the classifier performances strongly agree, so we can conclude that the SNN classifier is reliable in the context of the simulations used in this work.

\section{Method}\label{method}

In this section we describe the methodology employed to measure $f\sigma_8$ using the PVs derived from the SN samples described in Sect.~\ref{final_samples}. In this analysis we use the maximum likelihood method which assumes that the PV field is drawn from a multivariate Gaussian distribution. 
This method has been used in past works by \citet{johnson6dFGalaxyVelocity2014,adamsJointGrowthRate2020,laiUsingPeculiarVelocity2023,carreresGrowthrateMeasurementTypeIa2023} who applied the maximum likelihood methodology to samples of PVs derived from SNe Ia, TF, and FP galaxies. 

The likelihood is a multivariate Gaussian expressed as:
\begin{equation}\label{likelihood}
\begin{split}
        L (f\sigma_8,\Theta,\Theta_{\mathrm{HD}} )  =  (2\pi)^{-\frac{n}{2}} |C(f\sigma_8,\Theta,\Theta_{\mathrm{HD}} )|^{-\frac{1}{2}}   \\ 
        \times \exp \left[ -\frac{1}{2} \mathbf{v}^T(\Theta_{\mathrm{HD}}) C(f\sigma_8,\Theta,\Theta_{\mathrm{HD}} )^{-1} \mathbf{v}(\Theta_{\mathrm{HD}}) \right] \,,
\end{split}
\end{equation}
where $\mathbf{v}$ is the velocity data vector, $\Theta$ is the vector of the nuisance parameters, and $\Theta_{\mathrm{HD}}$ is the vector of the Hubble diagram (HD) standardization parameters. $C(f\sigma_8,\Theta,\Theta_{\mathrm{HD}})$ is the covariance matrix that describes the correlations between the PVs and includes the measurement errors.

In this section we present in detail all the components of the likelihood and the parameters of the model: in Sect.~\ref{PVestimation} we describe how we measure the PVs from the HD residual and in Sect.~\ref{covariance matrix} we explain the construction of the covariance matrix.

\subsection{Peculiar velocity estimation}\label{PVestimation}

The \sn\ host-PVs can be estimated using the residual of the HD after standardization. The \sn\ standardization process is the minimization of the residuals of the distance modulus $\mu_i$ to the HD, using the Tripp relation, given in Eq. \eqref{tripp}.
We define the parameter vector that contains the HD standardization parameters as $\Theta_{\mathrm{HD}} =  ( \alpha,   \beta,   M_0 )$.

To estimate the host-galaxy PVs we make use of the HD residuals, which are the difference between the observed distance modulus and the distance modulus given by the cosmological model and evaluated at the observed redshift of each SN. The residuals are given by
\begin{equation}
    \Delta \mu_i (\Theta_{\mathrm{HD}}) = \mu_{\mathrm{obs},i} (\Theta_{\mathrm{HD}}) - \mu_{\mathrm{model},i}(z_{\rm cos}) ,
\end{equation}
and their uncertainties
\begin{equation}
    \sigma^2_{\mu,i} (\Theta_{\mathrm{HD}},\sigma_M) =  A^T C_{\mathrm{SALT},i}  A + \sigma_M^2,
\end{equation}
where $C_{\mathrm{SALT},i}$ is the SALT covariance, defined in Eq. \eqref{salt_covariance}, $\sigma_M^2$ is the SN intrinsic scatter and $A^T = ( \alpha, -\beta, 1 )$. 

\citet{carreresGrowthrateMeasurementTypeIa2023}, following \citet{huiCorrelatedFluctuationsLuminosity2006}, demonstrates that the first-order expansion of HD residuals with respect to the PVs gives the following estimator:
\begin{equation}\label{PVestimator}
    v_i (\Theta_{\mathrm{HD}}) = - \frac{\ln(10) c} {5} \left( \frac{(1+z_{\mathrm{obs},i})c}{H(z_{\mathrm{obs},i}) r(z_{\mathrm{obs},i})} -1 \right) ^{-1}  \Delta \mu_i (\Theta_{\mathrm{HD}}).
\end{equation}

Naturally, the estimator gives the host-galaxy PV along the line of sight only. The uncertainty on the velocity estimation is:
\begin{equation}\label{PVerror}
    \begin{split}
        \sigma_{v,i} (\Theta_{\mathrm{HD}},\sigma_M) = \frac{\ln(10) c} {5} \left( \frac{(1+z_{\mathrm{obs},i})c}{H(z_{\mathrm{obs}s,i}) r(z_{\mathrm{obs},i})} -1 \right)  ^{-1} \\
        \times \sigma_{\mu,i} (\Theta_{\mathrm{HD}},\sigma_M).
   \end{split}
\end{equation}

\subsection{Covariance matrix}\label{covariance matrix}

The covariance matrix in Eq. \eqref{likelihood} is defined as:
\begin{equation}\label{fullvelcoveq}
    \begin{split}
    C_{ij}(f\sigma_8,\Theta,\Theta_{\mathrm{HD}}) &= C^{vv}_{ij} (f\sigma_8, \sigma_u) \\
    & +[\sigma^2_v + \sigma^2_{v,i}(\Theta_{\mathrm{HD}},\sigma_M)] \delta^K_{i,j}\,,
    \end{split}
\end{equation}
where $C^{vv}$ is the analytical part that depends on $f\sigma_8$ and $\sigma_u$, $\sigma^2_{v,i}$ is the error on the velocity estimator defined in Eq.\eqref{PVerror}, and $\sigma^2_v$ is a nuisance parameter to take into account the extra scatter due to the imperfect modeling of non-linear scales. Hereafter, we define the vector of the nuisance parameters as $\Theta = (\sigma_u, \sigma_v, \sigma_M)$.

The covariance $C^{vv}$ represents the correlation of the velocity field between the positions $\mathbf{r}_i$ and $\mathbf{r}_j$. In the general case we can define $C^{vv}_{ij} = \langle \mathbf{v_i} \mathbf{v}^*_j \rangle$, where $\mathbf{v}$ is the 3D velocity vector. The radial component $v_i$ of the 3D velocity field at a position $\mathbf{r}_i$ in Fourier space can be written as:
\begin{equation}
    v_i = \hat{\mathbf{r}_i} \cdot \mathbf{v}(\mathbf{r}_i) = \int \frac{d^3k}{(2\pi)^3} e ^{i\mathbf{k} \cdot \mathbf{r}_i} \hat{\mathbf{r}_i} \cdot \mathbf{v}(\mathbf{k}) \,.
\end{equation}

Therefore, the covariance can be defined as:
\begin{equation}\label{Cvv_def}
   C^{vv}_{ij} = \int \int \frac{\mathrm{d}^3k}{(2 \pi)^3} \frac{\mathrm{d}^3 k'}{(2 \pi)^3} e^{i(\mathbf{k\cdot r_i} - \mathbf{k' \cdot r_j})} \langle (\hat{\mathbf{r}}_i \cdot \mathbf{v}(\mathbf{k})) (\hat{\mathbf{r}}_j \cdot  \mathbf{v}^*(\mathbf{k}')) \rangle \,.
\end{equation}

We can rewrite Eq.\eqref{Cvv_def} using the velocity divergence scalar field, $\theta(\mathbf{r})$, defined as:
\begin{equation}
    \nabla \mathbf{v}(\mathbf{r},a) = - a f(a) H(a) \theta(\mathbf{r},a) \,,
\end{equation}
 where $a$ is the scale factor and $H(a)$ is the Hubble parameter. Assuming that the velocity field is irrational, in Fourier space we obtain:
 \begin{equation}
    \mathbf{v}(\mathbf{k},a)= -iaf(a)H(a) \theta(\mathbf{k},a) \frac{\hat{\mathbf{k}}}{k} \,. 
 \end{equation}
Using the divergence field we can define the velocity-divergence auto power spectrum as $\langle \theta(\mathbf{k}) \theta^*(\mathbf{k'})  \rangle = (2\pi)^3 \delta_D(\mathbf{k}-\mathbf{k}') P_{\theta\theta}(k)$.
At this point we can use the velocity-divergence auto power spectrum to simplify Eq.\eqref{Cvv_def}; the resulting $C^{vv}$ is given by: 
\begin{equation}\label{Cvv}
   C^{vv}_{ij} = \frac{(aHf)^2}{(2\pi)^2}\int_0^{+ \infty} P_{\theta\theta}(k) W_{ij}(k;\mathbf{r}_i,\mathbf{r}_j)dk,
\end{equation}
for the full derivation of the velocity covariance see \citet{johnson6dFGalaxyVelocity2014,carreresGrowthrateMeasurementTypeIa2023} and references therein.
In Eq. \eqref{Cvv}, $W_{ij}(k;\mathbf{r}_i,\mathbf{r}_j)$ is the so-called window function, which is defined as \citep[see][]{maPeculiarVelocityField2011}:

\begin{equation}\label{windowfunction}
    \begin{split}
        W_{ij}(k;\mathbf{r}_i,\mathbf{r}_j) & = \frac{1}{3} \left[ j_0 (k r_{ij}) -2 j_2 (k r_{ij}) \right] \cos(\phi_{ij}) \\
        & + \frac{1}{r_{ij}^2} j_2 (k r_{ij}) r_i r_j \sin^2(\phi_{ij}),
   \end{split}
\end{equation}
where $\phi_{ij}$ is the angle between $\mathbf{r}_i$ and $\mathbf{r}_j$,
$r_{ij}$ = |$\mathbf{r}_i$-$\mathbf{r}_j$|, $j_0$ and $j_2$ are the zeroth and second order spherical Bessel functions respectively.

In this work we compute $P_{\theta\theta}$ using the cosmological parameters of the Uchuu simulation, and we normalize the power spectrum for $\sigma_{8,fid}^2$. To compute $C^{vv}$ we use the empirical non linear model of $P_{\theta\theta}$ from \citet{belAccurateFittingFunctions2019}. This model is built by parametrizing the velocity-divergence power spectrum as $  P_{\theta\theta} = P_{\theta\theta,lin} \exp [-k (a_1 + a_2 k + a_3 k^2)]$, where $a_1$, $a_2$, $a_3$ are coefficients fitted in simulations and depend linearly on $\sigma_8$. $P_{\theta\theta,lin}$ is the linear velocity-divergence power spectrum, which is equal to the linear density-density power spectrum and is computed using the Boltzmann solver \texttt{CAMB}\footnote{\url{https://camb.info/}} \citep{lewis_camb_2000}. \citet{carreresGrowthrateMeasurementTypeIa2023} performed an accurate investigation on the impact of various $P_{\theta\theta}$ models on the $f\sigma_8$ measurement. From the results of that work, we conclude that the approximation from \citet{belAccurateFittingFunctions2019} is a good enough model for the velocity-divergence power spectrum.

To calculate $C^{vv}$ we compute the comoving distances $\mathbf{r}$, which are estimated using the observed redshifts, hence the distances are affected by PVs. Therefore, we have to take into account the RSD in our model for $C^{vv}$. We model the RSD using the empirical damping on the $P_{\theta\theta}$ small scales from \citet{kodaArePeculiarVelocity2014}. The damping is given by: 
\begin{equation}
    D_u(k) = \frac{\sin(k \sigma_u)}{(k \sigma_u)}\,,
\end{equation}
and is based on N-body simulations.

In this work $C^{vv}$ is computed at $z=0$ since the Uchuu mocks come from a snapshot at that redshift. Hence
\begin{equation}\label{cvv_finaleq}
    C^{vv}_{ij} = \frac{H_0^2}{2\pi^2} \frac{(f\sigma_8)^2}{(f\sigma_8)_{\mathrm{fid}}^2} \int_{k_{\mathrm{min}}}^{k_\mathrm{max}} f^2_\mathrm{fid} P_{\theta\theta}(k) D_u^2(k) W_{ij}(k;\mathbf{r}_i,\mathbf{r}_j) dk\,,
\end{equation}
where $f\sigma_{8,\mathrm{fid}}$ is the value of the growth rate from the Uchuu simulation and $f_\mathrm{fid} \simeq \Omega_m^{0.55}$. We highlight that with the maximum likelihood method we fit for the value of $\frac{(f\sigma_8)}{(f\sigma_8)_\mathrm{fid}}$. $k_\mathrm{min}$ and $k_\mathrm{max}$ are the minimum and maximum scales between which the power spectrum is computed. In the covariance computation we set $k_\mathrm{max} = 1$ \ihmpc\ to ensure convergence (see Appendix B in \citealt{carreresGrowthrateMeasurementTypeIa2023}) and $k_\mathrm{min}$ is imposed by the size of the N-body simulation box: $k_{\mathrm{min}}=2\pi/L_{\mathrm{box}}=4.6 \times 10^{-3}$ \hmpc.
To compute the covariance given by Eq. \eqref{cvv_finaleq} we use the \texttt{FLIP}\footnote{\url{https://github.com/corentinravoux/flip}} python package. The \texttt{FLIP} framework and the full derivation of the covariance matrix formula are presented in detail in \citet{ravoux2025flip}. In Appendix \ref{app:systematic_likelihood_fit} we show systematic tests which demonstrate the reliability of the maximum likelihood method.

\section{Results}\label{result}

In this section we describe the main results of this work.
Before measuring $f\sigma_8$ we check for any potential biases that could arise from our sample selection criteria (Sect.~\ref{selection_effect_HD_Vel}). After that, we use the maximum likelihood approach (Sect.~\ref{method}) to measure $f\sigma_8$ in different redshift bins for all the SN samples described in Sect.~\ref{final_samples}: see Sect.~\ref{resulst_full_sample} for the Full sample, Sect.~\ref{resulsts_specz_sample} for the Spec-z sample, and Sect.~\ref{result_contaminated_sample} for the Photo-typed sample. 


\subsection{Selection effect on the Hubble residuals and estimated velocities}\label{selection_effect_HD_Vel}

\citet{carreresGrowthrateMeasurementTypeIa2023} shows that the Malmquist bias heavily affects the growth-rate measurement through PVs since the likelihood in Eq.\eqref{likelihood} does not take into account selection effects.
As shown in Fig. \ref{fig:red_distr_all_samples}, at $z\sim 0.08$ all our SN samples start to be affected by the Malmquist bias. Therefore, to ensure that the final measurement will not be biased, we check the HD residuals and the estimated velocities as a function of redshift. In future works we will study how to correct for this type of selection bias using already existing method like BEAMS with Bias Corrections (BBC, \citealt{Kessler_2017_BBC}), UNITY \citep{Rubin_2015_unity,rubinUnionUNITYCosmology2023}, or directly modifying the likelihood as shown in \citet{Kim:2020_sampleselection}. 

\begin{figure}
    \centering
    \includegraphics[width=\hsize]{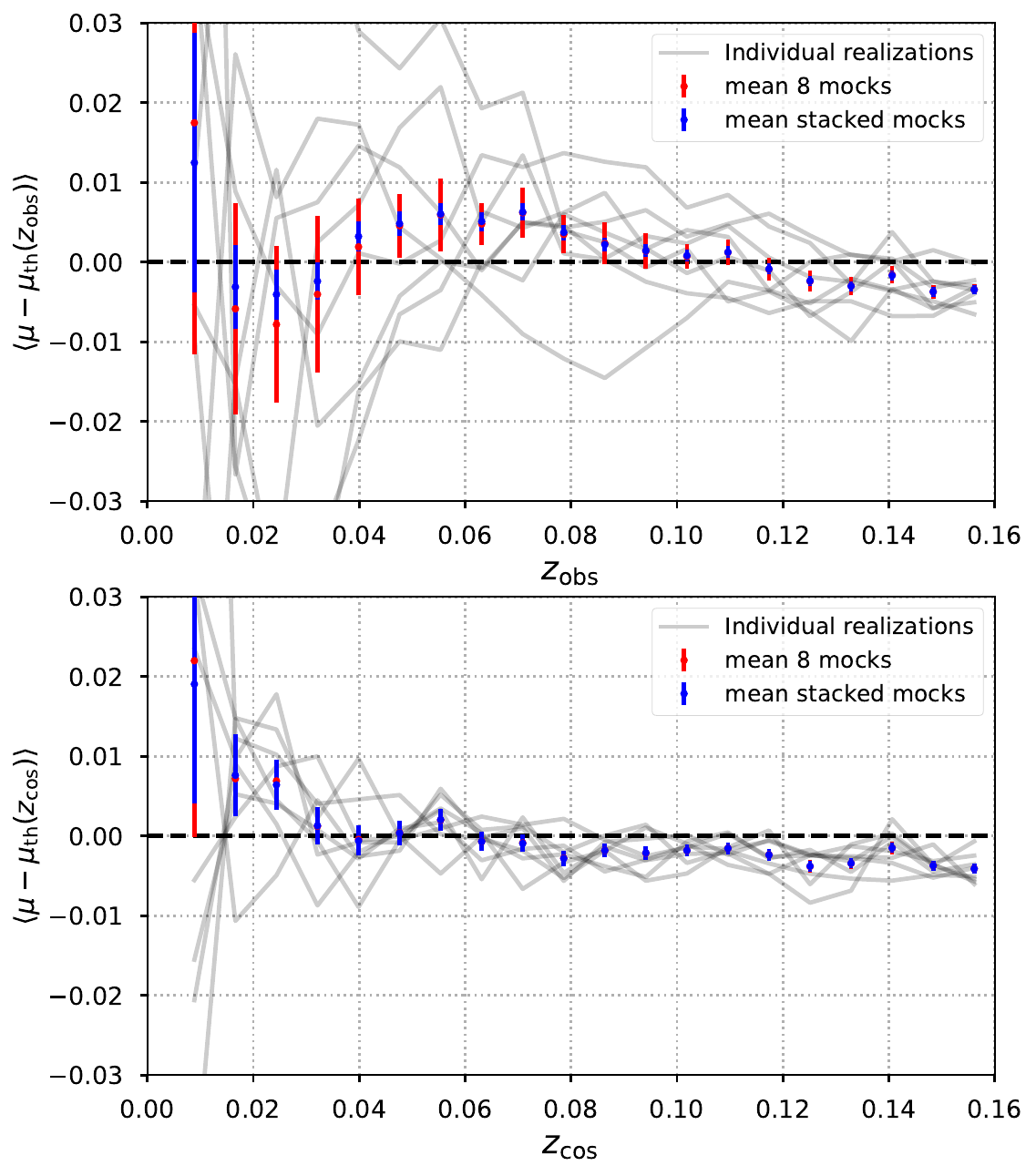}
    \caption{Hubble diagram residuals as a function of the observed redshift (top panel) and cosmological redshift (bottom panel) for the Full sample. The solid line grey lines show the residual for each of the $8$ LSST realizations, the red points show the mean of the 8 mocks, and the blue points show the mean computed by stacking all the mocks.}
    \label{fig:HD_residual_full_sample}
\end{figure}

Figure~\ref{fig:HD_residual_full_sample} shows the Full sample HD residuals as a function of the observed redshift (top panel) and as a function of the cosmological one (bottom panel). The HD residuals are computed using the input simulation values and the reference cosmology. We observe a trend in the HD residuals as a function of $z_{\mathrm{obs}}$ that follow the mean of the true galaxy PVs.
When the HD residuals are plotted against $z_{\mathrm{cos}}$, the trend disappears, indicating that PVs are responsible for the deviations in the residuals relative to $z_{\mathrm{obs}}$. 
The standard deviation of $8$ mocks (red points) is higher than the mean error over the stacking (blue points). This happens when the covariance due to PVs is not accounted for, as already shown in \citet{carreres2024ztfdr2}.
The discrepancy in the distance modulus residuals is evident at low redshift ($z<0.1$) and disappears at higher redshift, as the PV contribution becomes subdominant.
The bottom panel of Fig.\ref{fig:HD_residual_full_sample} shows a large reverse Malmquist bias for $z<0.02$, due to saturation effects, since the brighter objects are removed. 

\begin{figure}
    \centering
    \includegraphics[width=\hsize]{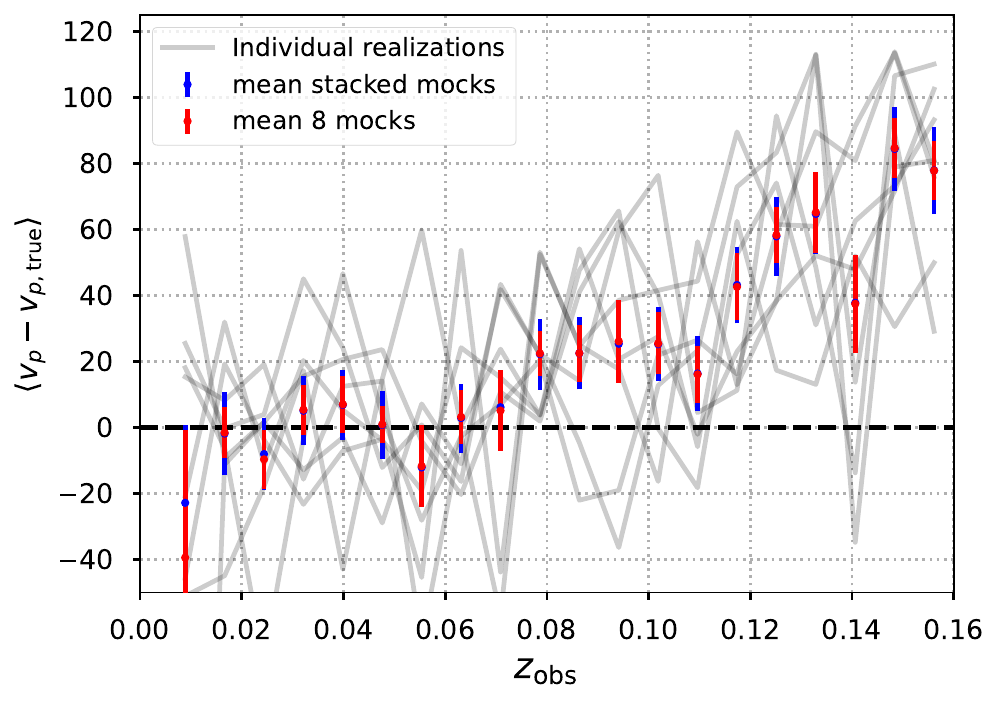}
    \caption{Difference between estimated PVs ($v_p$) and true PVs ($v_{p,\mathrm{true}}$) as a function of the observed redshift for the Full sample. The solid line grey lines shows the residual for each of the $8$ LSST realizations, the red points are the means over the 8 mocks, and the blue points show the means computed by stacking all the mocks.}
    \label{fig:velbias_full_sample}
\end{figure}

To ensure that the growth-rate measurement will not be biased we also check the PV residual. Figure~\ref{fig:velbias_full_sample} shows the difference between the PVs estimated from the HD residuals of Fig. \ref{fig:HD_residual_full_sample} and the true velocities as a function of the observed redshift. 
The PV residuals show a small bias for $z<0.02$ and $z>0.08$, the same range where the HD residuals are biased, as expected. The maximum PV bias is around $70$ \kms\,, which is almost one order of magnitude smaller than the PV measurement error in this redshift range. Therefore, we do not expect any significant bias on the growth-rate measurement. 

Considering the results on the PV bias, shown in Fig. \ref{fig:velbias_full_sample}, we choose to perform the growth-rate measurement within the redshift range $0.02<z<0.14$ for all the samples described in Sect.~\ref{final_samples}. 
The low-redshift cut is applied for two main reasons: to avoid the region where we have selection effects due to saturation and to avoid our PVs being correlated with the local structure nearby the observer, so we analyse only SNe which live inside the Hubble flow. The high redshift cut is applied to avoid the redshift range where the bias in PVs reaches about $100$ km s$^{-1}$.  
Figure~\ref{fig:HD_residual_full_sample} and Fig. \ref{fig:velbias_full_sample} show the Full sample residual plots only. We do not show
the HD and PV residuals for the Spec-z and Photo-typed sample since the results are almost identical to those of the Full sample.

\subsection{Full sample}\label{resulst_full_sample}

In this section we present the results of the growth-rate measurements from the PVs inferred from the Full sample, described in Sect.~\ref{SNIa_pure_sample_full}. 
We measure the growth-rate parameter in different redshift bins. We use volume bins to show the full statistical power of the LSST dataset going towards higher redshift. This means that for each redshift bin we use the same minimum redshift ($z_{\rm min}=0.02$) and increase the maximum redshift. We also measure the growth rate in tomographic bins to show the precision with which LSST PVs can explore the redshift dependence of $f\sigma_8$. When using the tomographic bins the correlations on the largest scales are not taken into account so the constraining power of the total sample decreases. Estimating $f\sigma_8$ as a function of redshift (in tomographic bins) allows us to measure the so-called "growth index" $\gamma$ \citep[see for example][]{Nguyen_2023}.

Figure~\ref{fig:fs8_result_all} and Table \ref{tab:results_all_sample} shows the fit results for each mock (top panel) and the relative error (bottom panel) for each sample. 

We focus on the left panel of Fig. \ref{fig:fs8_result_all}, which shows the results for the Full sample. We notice that the averages are always compatible with the fiducial value, except for the redshift range $0.06<z<0.1$. The discrepancy between the true $f\sigma_8$ and the recovered one in that redshift bin is less than 2-$\sigma$ for each realization. The small discrepancy is probably caused by sample variance. To confirm this, more than 8 realizations are needed, so
we leave this investigation for future works.
The mean reduced $\chi^2$ is about one in each bin, except in the redshift range $0.06<z<0.1$ where the mean reduced $\chi^2$ is $\sim 1.3$, which confirms the robustness of the fits. 
From the study of the optimistic Full sample we can appreciate the full statistical power of LSST data to constrain the growth rate using SN Ia PVs reaching an $8\%$ precision in the redshift range $0.02<z<0.14$. Moreover, in the most optimistic case we will be able to constrain the parameter to $12\%$ precision using tomographic bins.
The results described in this section demonstrate that the small bias in the PV residual, see Sect.~\ref{selection_effect_HD_Vel}, does not impact the growth-rate measurement. The small Malmquist bias does not impact the growth-rate measurements obtained with the Spec-z sample and the Photo-typed sample, as shown in the next sections. The Malmquist bias affects the fit results for the HD parameters, as described in Appendix \ref{app:fitHDpara}.

\begin{table}[t]
    \renewcommand{\arraystretch}{1.3}
    \centering
    \caption{ Summary of $f\sigma_8$ results for all LSST scenarios.}
        \begin{tabular}{cccrcc} 
        \hline 
        \hline 
        
            $z_{\mathrm{min}}$ & 
            $z_{\mathrm{max}}$ & 
            $\langle z_{\mathrm{obs}} \rangle$  & 
            $\langle N_{SN}\rangle$ & 
            $\left\langle \frac{f\sigma_8}{f\sigma_{8,\mathrm{fid}}} \right\rangle$ & $\left\langle \frac{\sigma_{f\sigma_8}}{f\sigma_{8,\mathrm{fid}}} \right\rangle$  \\ 
            
            \hline
            \multicolumn{6}{c}{Full sample} \\
            \hline
            
            $0.02$ &$0.06$ & $0.05$ &$2,969$ &$1.02$ &$0.14$ \\ 
            $0.02$ &$0.10$ & $0.08$ &$14,091$ &$0.97$ &$0.09$ \\
            $0.02$ &$0.14$ & $0.11$ &$36,599$ &$1.01$ &$0.08$ \\
            $0.06$ &$0.10$ & $0.08$ &$11,122$ &$0.93$ &$0.12$  \\
            $0.10$ &$0.14$ & $0.12$ &$22,507$ &$1.05$ &$0.12$  \\
            
            \hline
        
            \multicolumn{6}{c}{Spec-z sample} \\
            \hline
            
            $0.02$ &$0.06$ & $0.05$ &$1,989$ &$1.01$ &$0.16$ \\ 
            $0.02$ &$0.10$ & $0.08$ &$9,500$ &$0.99$ &$0.12$ \\
            $0.02$ &$0.14$ & $0.11$ &$24,194$ &$1.01$ &$0.09$ \\
            $0.06$ &$0.10$ & $0.08$ &$7,511$ &$0.94$ &$0.14$  \\
            $0.10$ &$0.14$ & $0.12$ &$14,693$ &$1.01$ &$0.15$  \\
            
            \hline
            \multicolumn{6}{c}{Photo-typed sample} \\
            \hline
            
            $0.02$ &$0.06$ & $0.05$ &$1,917$ &$1.01$ &$0.18$ \\ 
            $0.02$ &$0.10$ & $0.08$ &$9,345$ &$0.99$ &$0.12$ \\
            $0.02$ &$0.14$ & $0.11$ &$23,903$ &$1.02$ &$0.10$  \\
            $0.06$ &$0.10$ & $0.08$ &$7,428$ &$0.93$ &$0.14$  \\
            $0.10$ &$0.14$ & $0.12$ &$14,558$ &$1.02$ &$0.15$  \\
        \end{tabular}
    \label{tab:results_all_sample}
\end{table}

\subsection{Spec-z sample}\label{resulsts_specz_sample}

\begin{figure*}
    \centering
    \includegraphics[width=\hsize]{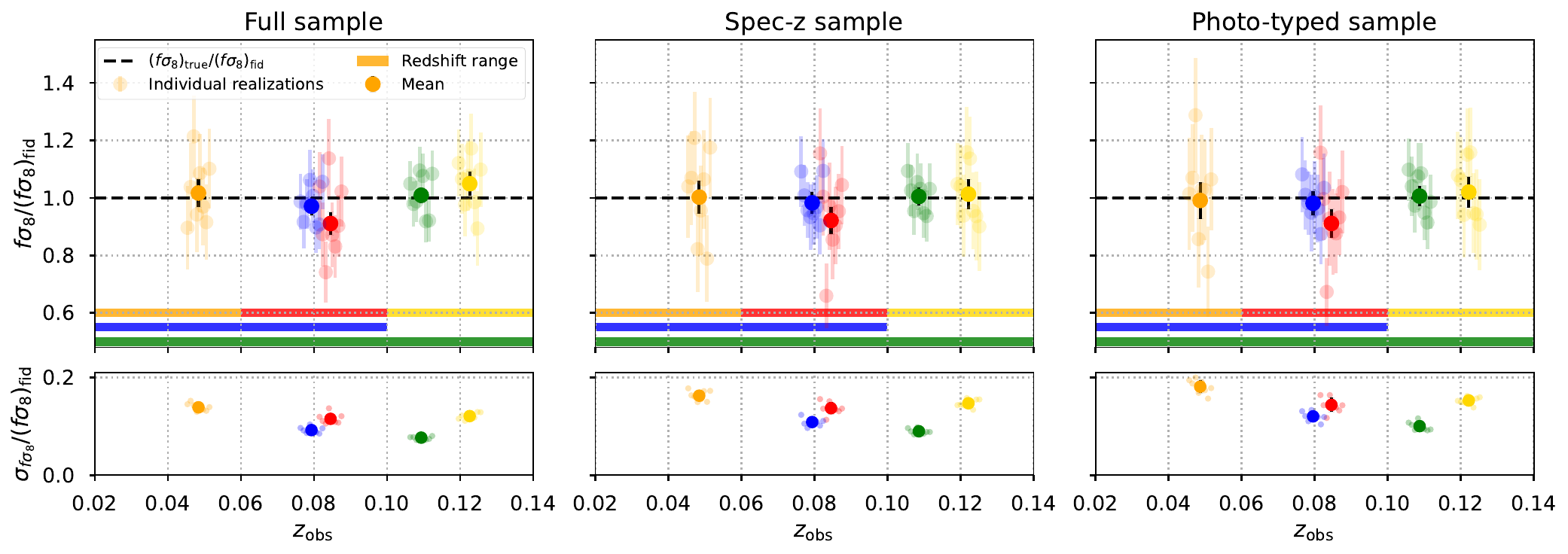}
    \caption{Results of $f\sigma_8$ measurements for all the scenarios in different redshift bins. \textit{Left:} Full sample, \textit{centre:} Spec-z sample, and \textit{right:} Photo-typed sample. For each scenario, the top panel shows the results of the fit for each of the 8 LSST realizations together with the mean value, while the bottom panel shows the mean values of $\sigma_{f\sigma_8}/f\sigma_{8,fid}$, which are the relative errors on the growth-rate measurements. The averages are plotted at the mean redshift positions, while the results for the individual realizations are slightly shifted along the abscissa to improve the clarity of the figure. The colored bars show the redshift range used for each $f\sigma_8$ measurement; the bars have the same colors as the measurements.}
    \label{fig:fs8_result_all}
\end{figure*}

We present the growth-rate measurement using the PVs measured from the Spec-z sample, described in Sect.~\ref{SNIa_pure_sample_spectro_selected}. 
Figure~\ref{fig:fs8_result_all} (central panel) and Table \ref{tab:results_all_sample} show the fit results for the Spec-z sample. The recovered $f\sigma_8$ are always compatible with the simulation input, except for the redshift bin $0.06<z<0.1$ as in the previous case. For the Spec-z sample the mean reduced $\chi^2$ is about one in each bin, except in the redshift range $0.06<z<0.1$ where the mean reduced $\chi^2$ is $\sim 1.2$, which confirms the goodness of the fits. 
From the study of the Spec-z sample we can see that LSST, using the spectroscopic redshift from DESI and 4MOST, will be able to constrain the growth rate with $9\%$ precision in the redshift range $0.02<z<0.14$. Moreover, in this more realistic scenario the constraints on $f\sigma_8$ reach $14\%$ precision in the tomographic bins.

\subsection{Photo-typed sample}\label{result_contaminated_sample}

In this section we describe the growth-rate measurement using the PVs measured from the Photo-typed sample described in Sect.~\ref{contaminated_sample}. This is our baseline analysis and the most realistic LSST scenario. 

The results for the growth-rate measurement using the Photo-typed sample are shown in Figure~\ref{fig:fs8_result_all} (right panel) Table \ref{tab:results_all_sample}. Figure~\ref{fig:fs8_result_all} shows that the averages are always compatible with the simulation input, except for the redshift bin $0.06<z<0.1$ as in the previous cases.
From the Figure~we can conclude that in this scenario there is also no significant bias in the measurement and we are able to recover correctly the growth rate. Therefore, a low percentage of contamination in the HD does not bias $f\sigma_8$ from PVs. 

We note that in this scenario the error on $f\sigma_8$ is systematically higher ($1-2 \%$) compared to the one for the Spec-z sample. This is caused by the slightly lower number of SNe Ia in the Photo-typed sample due to SNN inefficiencies, and by the extra scatter in the HD due to the contamination. To test this we perform two extra fits: one by removing the contaminants and the other by manually adding SNe Ia to reach the same numbers as in the Spec-z sample. In both cases the average error of $f\sigma_8$ is higher with respect to the one in the Spec-z sample, but not as large as in the Photo-typed sample. This test demonstrates that a low level of contamination does not bias the measurement but slightly increases the final error. In the next Section we show the effect of increasing contamination in the measurement of the growth-rate.
Additionally, the mean reduced $\chi^2$ for the Photo-typed sample is $\sim 1.05$ in each bin and $\sim 1.2$ in the redshift range $0.06<z<0.1$. These values demonstrate the goodness of the fits and are comparable to the ones of the Spec-z sample.
From the study of the Photo-typed sample we can see that LSST will be able to constrain the growth rate with $10\%$ precision in the redshift range $0.02<z<0.14$. Moreover, we can constrain the $f\sigma_8$ up to $14\%$ precision in the tomographic bins.

In Appendix \ref{forecast_across_year}, using the Photo-typed sample, we show forecasts for the expected error on the $f\sigma_8$ across the LSST survey years of observation. We demonstrate that after 5 years of observation, we will be able to reach a precision of $12\%$ in the redshift range $0.02<z<0.14$.

\section{Test of the effect of contamination}\label{bias_contamination}

In this Section, we vary the sample contamination to determine the threshold at which the incorrect likelihood gives significantly biased results. 
To produce this variation we do not use SNN. In each new sample we use the SNe Ia inside the Spec-z sample and we vary the SN contamination\footnote{The contamination is defined as the number of contaminants divided by the total number of SNe in the sample. This is the most natural definition for simulated objects as we know their true type.} by applying different quality cuts on the SALT fit results of the other SN types \citep[as similarly done in][]{vincenzi_2021_descontamination}.

\begin{figure}
    \centering
    \includegraphics[scale=0.7]{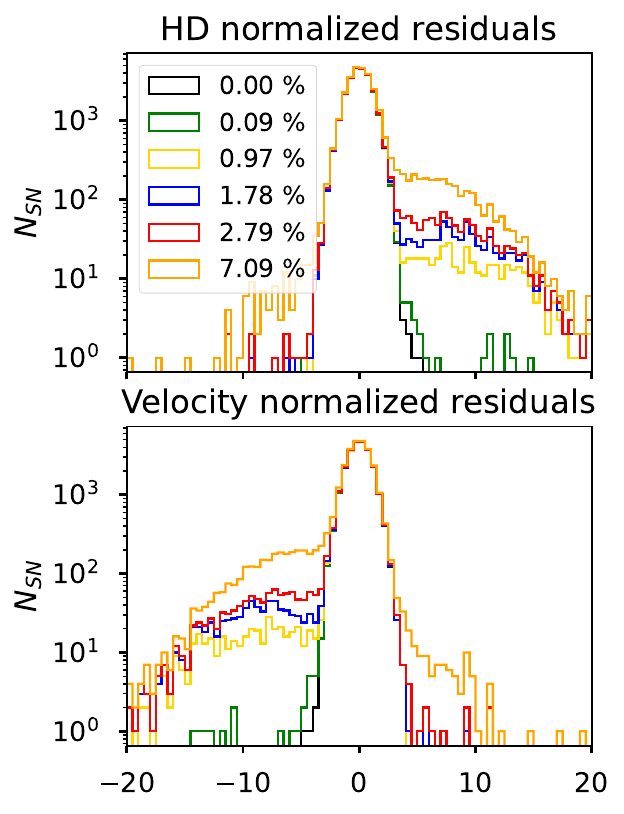}
    \caption{HD residuals (top panel) and the residuals of the estimated velocity (bottom panel), normalized by the errors, using the input HD parameters of the simulation. The plot shows the residuals for a random realization in the simulation in the redshift range $0.02<z<0.14$. The different colors represent different SN samples with different levels of contamination, as shown in the legend.  }
    \label{fig:residual_contaminated_salt}
\end{figure}

Figure~\ref{fig:residual_contaminated_salt} illustrates the HD residuals and the residuals of the estimated velocities, normalized by the errors, for different samples with increasing contamination for one LSST realization. The HD residual and the velocity pulls are computed using the input value of the simulation in the redshift range $0.02<z<0.14$. Figure~\ref{fig:residual_contaminated_salt} shows that as the contamination increases, the distributions become more skewed. This can create a bias in the maximum likelihood, since the incorrect likelihood describes a Gaussian velocity distribution. To study when the method breaks, we perform the $f\sigma_8$ fit on the samples with different contamination.

\begin{figure}
    \centering
    \includegraphics[width=\hsize]{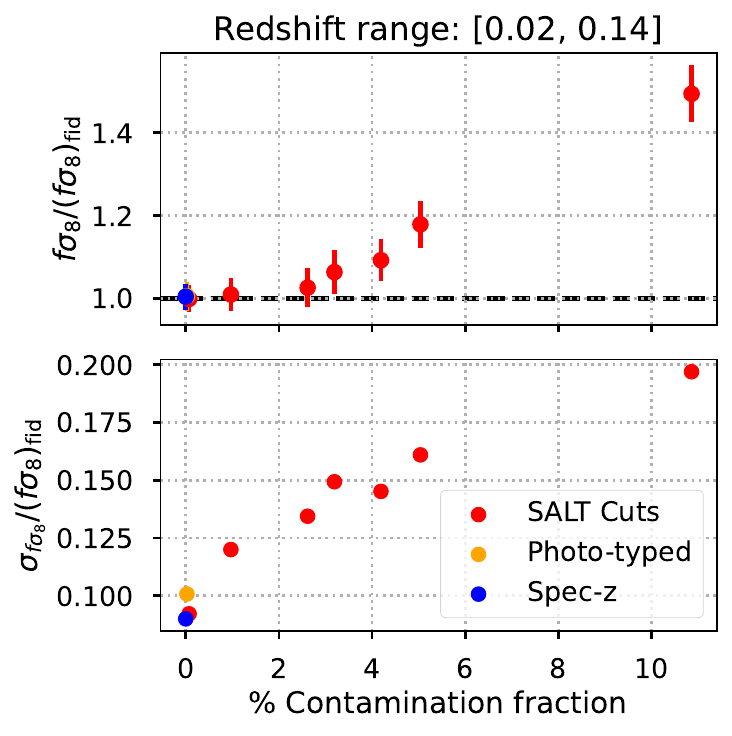}
    \caption{$f\sigma_8$ fit results (top panel) and relative error (bottom panel) as a function of the sample contamination fraction for the redshift bin $0.02<z<0.14$. The blue points show the results for the Spec-z sample, the orange ones for the Photo-typed sample, and the red ones for the different samples created using the cuts on the SALT fits as classification (SALT cuts). Every point is the average over the 8 LSST realizations.}
    \label{fig:fs8_result_contaminated_salt}
\end{figure}

Figure~\ref{fig:fs8_result_contaminated_salt} presents the $f\sigma_8$ results, averaged over the 8~LSST realizations, as a function of contamination, for the redshift bin $0.02<z<0.14$. 
As the level of contamination increases, the bias on the growth-rate measurement becomes larger, especially when including higher redshift SNe. The measurement error on  $f\sigma_8$ also increases as a function of the contamination in the sample, because of the higher scatter in the HD diagram, as explained in the previous section. Figure~\ref{fig:fs8_result_contaminated_salt} shows that a contamination level above~$\sim 2 \%$ breaks the likelihood model. As the contamination increases, the HD parameters become more biased. We highlight that for high contamination fraction (i.e. $> 2 \%$) the fit fails in some realizations and gives $f\sigma_8 \sim 0$. This is because the scatter in the HD diagram increases to the point where the information of the PV correlation is lost. We find the same results for the other redshift bins listed in Table \ref{tab:results_all_sample}.
Additionally, we observe that the measurement error for the red point with the lowest contamination is smaller than the error with the Photo-typed sample (i.e. SNN), but still higher compared to the error of the pure SNe Ia. As explained in Sect. \ref{result_contaminated_sample}, the number of SNe Ia in the Photo-typed sample decreases, due to the efficiency of the SNN.
The results shown in Fig. \ref{fig:fs8_result_contaminated_salt} emphasize the need to control contamination to ensure the accuracy of the cosmological parameter estimation.

\section{Discussion}\label{discussion}

\begin{figure}
    \centering
    \includegraphics[width=\hsize]{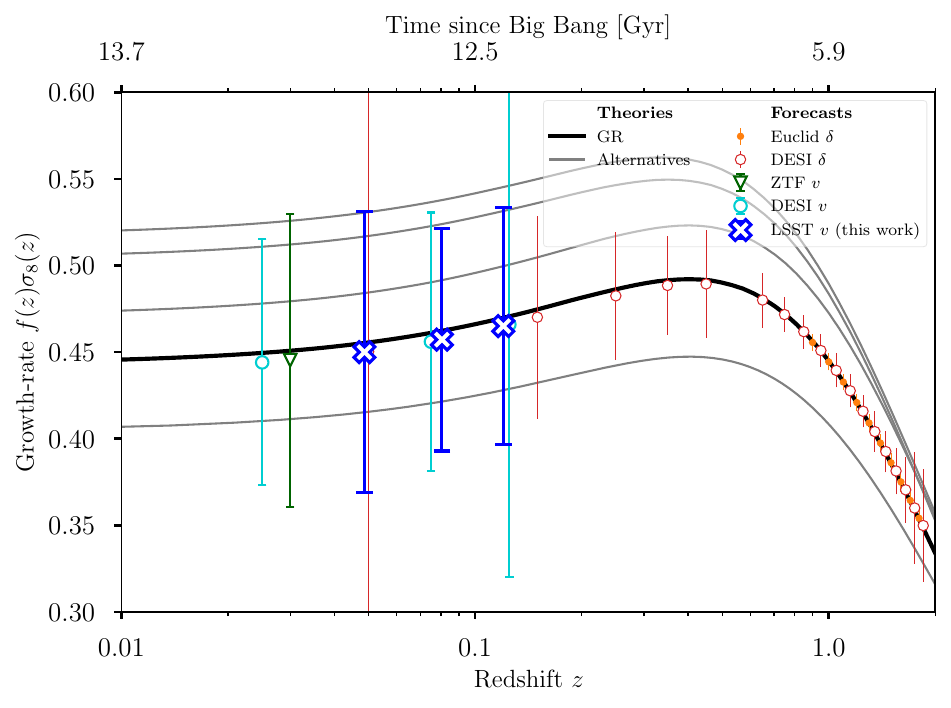}
    \caption{Growth-rate parameter as a function of redshift. The black solid line shows the GR-$\Lambda$CDM prediction, while the gray lines present the predictions for alternative gravity theories. The points show different forecasts for future and current stage IV surveys. The orange points show the forecast for Euclid \citep{amendolaFateLargescaleStructure2018}, the red circles shows the ones for DESI \citep{desicollaboration2016desi}. The forecasts from DESI and Euclid are derived from the analysis of the density field. The other forecasts come from the PV field: the green triangle from ZTF SNe Ia \citep{carreresGrowthrateMeasurementTypeIa2023}, the cyan circles from DESI PV survey \citep{saulderTargetSelectionDESI2023}, and the blue crosses from LSST SNe Ia, which are the results from the Photo-typed sample in the tomographic bins. }
    \label{fig:fs8_redshift_survey}
\end{figure}

Figure~\ref{fig:fs8_redshift_survey} illustrates the predicted evolution of the growth-rate parameter $f\sigma_8$ as a function of redshift, comparing predictions from both General Relativity (GR-$\Lambda$CDM) and alternative gravity theories. The Figure~also shows forecasts for current and upcoming stage IV surveys like Euclid \citep{amendolaFateLargescaleStructure2018}, DESI \citep{desicollaboration2016desi,saulderTargetSelectionDESI2023}, and ZTF \citep{carreresGrowthrateMeasurementTypeIa2023}.The blue crosses represent the LSST PVs forecasts estimated in this analysis from the Photo-typed sample for the tomographic redshift bins. 
When compared with the results from ZTF PVs and DESI PVs, the forecast from LSST shows similar results for $z<0.06$. LSST will obtain a precision similar to that of ZTF since both surveys will have a complete sample up to $z<0.06$ with a similar number of SNe inside a similar volume. However, Fig. \ref{fig:fs8_redshift_survey} also shows that LSST will obtain tighter constraints compared to DESI and ZTF for $z>0.06$, and will therefore be able to more precisely measure the evolution of the growth-rate parameter with respect to the other surveys.

When compared to the forecast from \citet{howlett_LSST_forecast} ($>10\%$ for $z>0.1$), the forecasts from this work give larger uncertainties. This can be explained by the fact that in \citet{howlett_LSST_forecast} the authors produced an overly idealistic simulation for LSST, without passing through all the selection steps described in Sect.~\ref{selection_DIA}. Therefore, in that study the authors use a larger number of SNe Ia with respect to this work. Moreover, in \citet{howlett_LSST_forecast} the authors produced their results using the Fisher forecast method which usually underestimates the error and does not take into account the geometry of the survey footprint. In contrast, in this work we use the maximum likelihood and we fit both $f\sigma_8$, the HD parameters, and the nuisance parameters all together, as real data should be analyzed. Our results are more realistic with respect to the ones in \citet{howlett_LSST_forecast} since our simulations are more realistic (including photometric classification) and we use the maximum likelihood method to extract the growth-rate constraints.

\section{Conclusions}\label{conclusion}

This work demonstrates that the LSST SN sample can be a powerful dataset for measuring the growth rate of cosmic structures, $f\sigma_8$, through PVs. We simulate LSST light curves over 10 years for various types of SNe (Sect.~\ref{lsstsim}) using a realistic observing  strategy, incorporating noise and the effect of the PV field coming from the Uchuu simulation. We create 8 different realizations of LSST 10-year survey using the detection efficiency of the DIA pipeline (Sect.~\ref{selection_DIA}) and the host spectroscopic redshift efficiency coming from DESI and 4MOST (Sect.~\ref{host_spectra_selection}).

We explore three scenarios: an idealized fully spectroscopic redshift sample, a spectroscopic sample combining LSST with DESI and 4MOST to obtain the SN host spectroscopic redshift, and a photometric sample incorporating machine-learning-based classification (Sect.~\ref{photometric_classification}). These scenarios have been built to assess the potential of LSST under various conditions and disentangle possible systematics. For every scenario we use the maximum likelihood method (Sect.~\ref{method}) to obtain the growth-rate measurement using the PVs measured from the SN HD residuals.

In the most optimistic scenario, the Full sample, containing $\langle 52,326 \rangle$ SNe Ia up to $z=0.16$, we achieve the highest statistical precision for $f\sigma_8$. The results demonstrated a uniform sky distribution within the WFD area (see Fig.~\ref{fig:density_full_sample}) and a nearly flat comoving density at low redshift, as shown in Fig. \ref{fig:red_distr_all_samples}. The Full sample, as well as the other ones, suffers from a small Malmquist bias, which does not impact the final constraints on the growth rate (Sect.~\ref{selection_effect_HD_Vel}).
For this sample, the measurement of $f\sigma_8$ closely matches the input value of the simulation with minimal statistical error: $8\%$ precision in the redshift range $0.02<z<0.14$ and $12\%$ precision for the tomographic bins (Sect.~\ref{resulst_full_sample}).

The Spec-z sample, incorporating spectroscopic redshift efficiencies from DESI and 4MOST, reduces the sample size to $\langle 33,682 \rangle$ SNe Ia due to a limited overlapping sky coverage (Fig. \ref{fig:density_specz_sample}). This sample maintains a high degree of uniformity within the overlapping survey footprints, and the comoving density was largely unaffected up to $z=0.12$, as shown in Fig. \ref{fig:red_distr_all_samples}. This scenario represents a realistic expectation of LSST synergy with other spectroscopic surveys, yielding competitive constraints on $f\sigma_8$ with a modest increase of the statistical uncertainties compared to the Full sample, see Sect.~\ref{resulsts_specz_sample}.

In the Photo-typed sample, we account for photometric classification using the SNN framework. Despite challenges such as sample contamination and efficiency limitations, the contamination fraction was successfully controlled at the level of $0.02\%$ (Sect.~\ref{contaminated_sample}). This indicates the robustness of SNN in handling simulated LSST data. The resulting $f\sigma_8$ measurements, while slightly degraded compared to the spectroscopic samples, still provides valuable constraints on the growth rate and shows LSST capability for independent cosmological analysis without relying heavily on a spectroscopic follow-up of each SN~Ia. After 10 years of LSST, we will be able to constrain the growth-rate parameter at the $10\%$ level using SN PVs only. Moreover, we will be able to measure $f\sigma_8$ reaching a $14\%$ precision in tomographic bins.


We also investigate the effect of contamination when using the likelihood of Eq.\eqref{likelihood}, see Sect.~\ref{bias_contamination}. We demonstrate that the method produces unbiased results, but with higher uncertainties, when the contamination is below $\sim 2\%$.

Overall, this work highlights the potential of the LSST PV sample to complement RSD measurements, offering a robust and independent way for testing GR and different dark energy models. The combination between high and low redshift $f\sigma_8$ measurements from the current generation surveys opens the door to more precise measurement of the "growth-index" $\gamma$ \citep{Nguyen_2023}.
Future studies should focus on constructing a model-motivated likelihood to include selection effects, HD residual correlation with host galaxies and SN colors, and to incorporate the probability coming from the photometric classifier. These efforts will enhance the LSST contribution to precision cosmology. To further improve the constraints on the growth-rate at low redshift it is possible to combine LSST PVs with other PV data from ZTF or DESI and to cross-correlate the LSST PV field with the density field coming from spectroscopic surveys (e.g. 4MOST and DESI).

\begin{acknowledgements} \\

This paper has undergone internal review in the LSST Dark Energy Science Collaboration. We thanks the internal reviewers: Rebecca Chen and Erin Hayes. 
\textbf{Author contributions:} D. Rosselli led the project, developed the simulation framework, performed the analysis, and wrote the manuscript.
B. Carreres contributed to the development and validation of the galaxy mocks. C. Ravoux contributed to the determination of the host spectroscopic redshift efficiencies. C. Ravoux, B. Carreres, and A.G. Kim contributed to the velocity covariance modeling.
J. E. Bautista provided support with plotting and reviewed the manuscript. D. Fouchez, F. Feinstein, B. Sánchez, and A. Valade contributed through manuscript review and critical feedback. B. Sánchez contributed to the light curve selection method and photometric classification task.
B. Racine contributed to the analysis of the simulations. 
All authors contributed to the critical discussion of the results and reviewed the manuscript. 
This work was conducted within the LSST Dark Energy Science Collaboration, which provided the infrastructure, tools, and coordination essential to this study.

\textbf{Grants:} The project leading to this publication has received funding from Excellence Initiative of Aix-Marseille University - A*MIDEX, a French ``Investissements d'Avenir'' program (AMX-20-CE-02 - DARKUNI). 
This work has been carried out thanks to the support of the DEEPDIP ANR project (ANR-19-CE31-0023).
This work received support from the French government under the France 2030 investment plan, as part of the Initiative d'Excellence d'Aix-Marseille Université - A*MIDEX (AMX-19-IET-008 - IPhU).
This work was performed using the Dark Energy Center (DEC) hosted at Aix Marseille Univ, CNRS/IN2P3, CPPM, Marseille, France.
The DESC acknowledges ongoing support from the Institut National de Physique Nucl\'eaire et de Physique des Particules in France; the Science \& Technology Facilities Council in the United Kingdom; and the Department of Energy and the LSST Discovery Alliance in the United States.  DESC uses resources of the IN2P3 Computing Center (CC-IN2P3--Lyon/Villeurbanne - France) funded by the Centre National de la Recherche Scientifique; the National Energy Research Scientific Computing Center, a DOE Office of Science User 
Facility supported by the Office of Science of the U.S.\ Department of Energy under Contract No.\ DE-AC02-05CH11231; STFC DiRAC HPC Facilities, funded by UK BEIS National E-infrastructure capital grants; and the UK particle physics grid, supported by the GridPP Collaboration. This 
work was performed in part under DOE Contract DE-C02-76SF00515.
\textbf{Codes:} We acknowledge the use of \texttt{iminuit} (\citealt{dembisky_iminuit_2020}), \texttt{numpy} (\citealt{Harris_2020}), \texttt{scipy} (\citealt{Virtanen_2020}), and \texttt{healpix} (\citealt{Gorski_2005}).
\textbf{Data availability:} The data that support the findings of this study are available from the corresponding author upon reasonable request. Access to the data may be granted to researchers who are not LSST collaboration members, subject to applicable data sharing policies and guidelines.

\end{acknowledgements}

\bibliographystyle{aa}
\bibliography{MyBib.bib,newBIB.bib}

\begin{appendix}

\section{Details on the computation of the SN/host spectroscopic redshift efficiencies}\label{app:r_host_details}

In this Section we describe the details of the computation of the Sn/host spectroscopic redshift efficiencies (see Sect. \ref{host_spectra_selection}. Following Eq.\eqref{eq:Rhost}, we need to compute the terms $R_{spectro}(z)$ and $R_{\rm SN}(z)$.

To determine $R_{spectro}$ we apply magnitude and color cuts to select the galaxies that will be targeted by DESI BGS sample and 4MOST for the CRS-BG and 4HS surveys. 

For the DESI BGS survey, we apply the main $r < 20.175$ magnitude limit selection cuts following \citet{hahn_desi_2023}. We do not include the fiber magnitude cut as the DESI-like fiber magnitude is not included in the cosmoDC2 simulation, and this cut is mainly there to reject imaging artefacts and shredded galaxies that are not considered in cosmoDC2. We do not apply the other BGS quality cuts for the same reason.

Following the descriptions in \citet{4most_crs_bg} for the 4MOST CRS-BG sample we apply the magnitude cuts $J < 16$, $K < 18$, and the color cuts $-1.6 < (J - K) - 1.6(J- W_{1}) < -0.5$, $(J - K) + 2.5(J- W_{1}) > 0.1$ and $(J - K) + 0.5(J- W_{1}) > 0.1$. For the 4MOST 4HS sample, following \citet{4most_4hs}, we apply the cuts $J < 18$ and $(J - K) < 0.45$.
Here, we assume that the Roman magnitudes in the cosmoDC2 catalog are close enough to WISE and SDSS magnitudes. This is not a strong assumption for the \textit{J, K} and \textit{r} magnitudes, but is for the $W_{1}$ magnitude which differs from the K213 Roman magnitude. However, this assumption has a minor effect compared to the main magnitude cuts. The main target selection is performed using the magnitude cuts, a small shift in the color does not change significantly the galaxy population in the considered redshift range. We run some tests by changing the color cuts and we find that $R_{\rm host}$ is not impacted.

To determine $R_{\rm SN}$ we simulate each SN type inside the cosmoDC2 simulation. To simulate the SN sample, we use the rates described in Sect.~\ref{snesim}. We model the correlation between SN Ia rates and galaxy properties following the "A + B" model from \citet{Mannucci_A+B}. The SN Ia rate is described as:
\begin{equation}\label{eq:mannucci_a+b}
     R^{\mathrm{A+B}}_{\mathrm{Ia}} (M_*,SFR) = A \times M_* + B \times SFR,
\end{equation}
where $M_*$ is the galaxy stellar mass and $SFR$ is the star formation rate. We use the best-fitting A and B parameters from \citet{sullivan_sniahost_2006}. To model the correlation for the other types of SNe we use the same approach described in \citet{vincenzi_2021_descontamination}. The correlation for peculiar SNe Ia (91bg-like and SN Iax) is modeled as the one of SNe Ia, but we set their rate to zero in passive galaxies. Passive galaxies a have specific star formation rate ($sSFR$) smaller than $10^{-11.5}$ yr$^{-1}$. SNe core-collapse occur almost exclusively in star-forming galaxies \citep{LI_2011}; following \citet{vincenzi_2021_descontamination}, we model this correlation as:
\begin{align*}
     &R_{\mathrm{Ib/c, II}} = 0 \mbox{ in passive galaxies}\\
     &R_{\mathrm{Ib/c}} \propto M_*^{0.36} \\
     &R_{\mathrm{II}} \propto M_*^{0.16}.
 \end{align*}

Once we have assigned the SNe to the host galaxy, we use the synthetic magnitudes from the cosmoDC2 simulation to determine which galaxies will be observed by the spectroscopic surveys using the same cuts as for $R_{spectro}$.

\section{Test of the robustness of the photometric classification}\label{app:SNNvsparsnip}

In this appendix we show the comparison between SNN and Parsnip\footnote{\url{https://github.com/LSSTDESC/parsnip}} \citep{Boone_2021_parsnip} to test the robustness of the photometric classification. We run the Parsnip classifier using the same training sample used for SNN and defined in Sect.~\ref{photometric_classification}.

We compare the two classifiers on the 8 LSST realizations. The following values are the mean and the standard deviations of these 8~realizations.
From Parsnip classification we initially identify a sample of $71,320$ SN candidates as likely SNe Ia, using the probability cut $P_{Ia}>0.5$. We find a classification efficiency of $96.72 \pm 0.06\%$, with contamination fraction at $1.09 \pm 0.03\%$. For SNN we have $68,084$ SNe Ia candidates with $92.3 \pm  0.1\%$, with contamination fraction at $1.15 \pm 0.04\%$.

\begin{figure}
     \centering
      \includegraphics[width=\hsize]{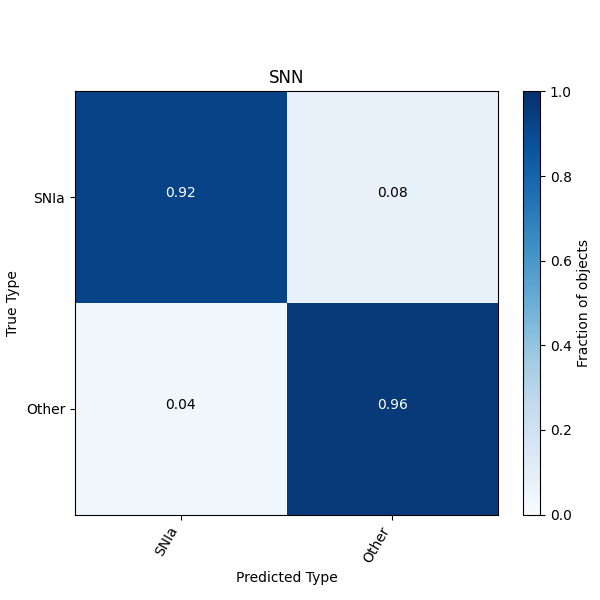}
      \caption{Confusion matrix for SNN classification for one LSST realization.}
      \label{fig:conf_matrix_SNN}
\end{figure}

\begin{figure}
      \centering
     \includegraphics[width=\hsize]{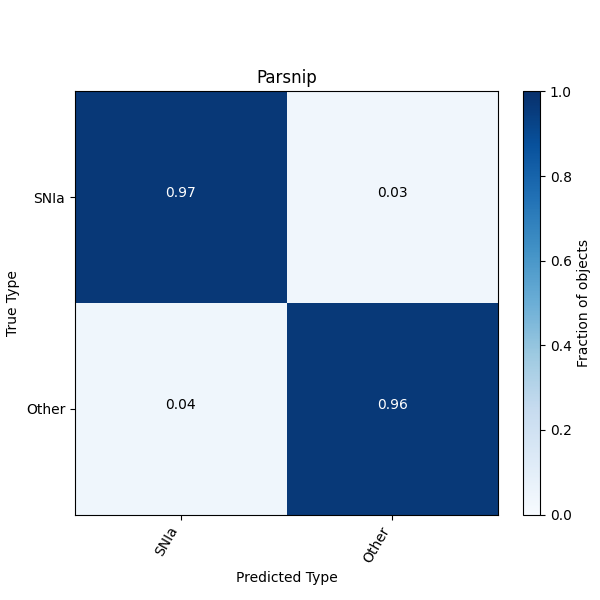}
      \caption{Confusion matrix for Parsnip classification for the same LSST realizations as Fig. \ref{fig:conf_matrix_SNN}.}
      \label{fig:conf_matrix_parsnip}
\end{figure}

Parsnip is more efficient compared to SNN, while the level of contamination in the sample is comparable. Figure~\ref{fig:conf_matrix_SNN} and Figure~\ref{fig:conf_matrix_parsnip} show the confusion matrices from SNN and Parsnip classification for one LSST realization respectively. The confusion matrices show the efficiency difference on the first row, while the second row is the same for both classifier (same contamination fraction). The higher efficiency of Parsnip is most probably related to the fact that Parsnip uses the information of the intrinsic magnitude of the objects to perform the classification, while SNN does not use that feature.

By imposing the additional quality selection through SALT fit
cuts (Sect.~\ref{SNefit}), the Parsnip sample is refined to $33,531$ SNe with $0.032 \pm 0.004 \%$ contamination fraction. For SNN we have $33,269$ SNe with $0.021 \pm 0.007 \%$ contamination fraction. The final contamination is slightly lower in the SNN classified sample. The difference between the two samples is really small and not significant for the growth-rate results described in Sect.~\ref{result_contaminated_sample}. From the comparison of the two different classifiers we can conclude that the photometric classification is robust, the classifiers give comparable results (except for efficiency). We highlight that SNN performances can be improved by using a larger training sample (N$>100,000$) \citep{Moller_2019_SNN,Vincenzi_2022}.

            
    

\section{Test of the likelihood method}\label{app:systematic_likelihood_fit}

\begin{figure}
    \centering
    \includegraphics[width=\hsize]{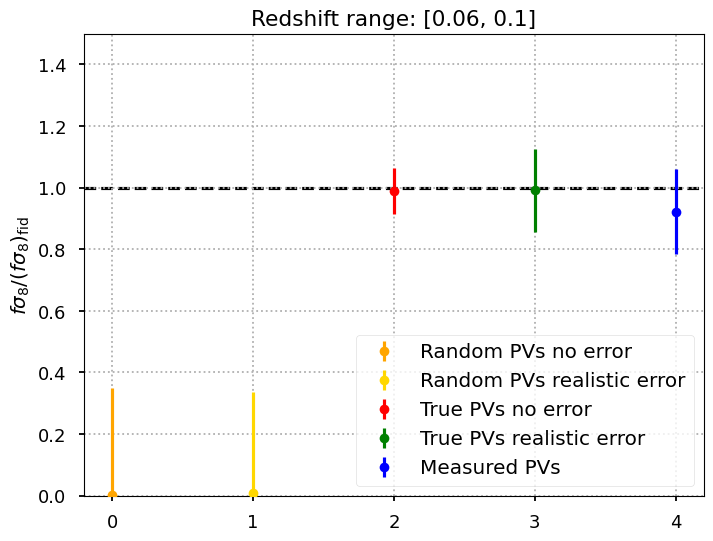}
    \caption{$f\sigma_8$ results in the redshift range $0.06<z<0.1$ for five different data vectors as defined in the legend. The colored points shows the averages over the 8 LSST realizations with the average relative errors.}
    \label{fig:fs8_test_systematics}
\end{figure}

In this section we describe the tests to check the reliability of the maximum likelihood method. 
Firstly, we test the method by changing the data vector (i.e. PVs). We perform different fits using the same SNe Ia but assigning different PVs. The results of these tests are shown in Figure~\ref{fig:fs8_test_systematics} in the case of the Spec-z sample in the redshift range $0.06<z<0.1$, the results remain the same for all the redshift bins used in Sect.~\ref{result}. 
The Figure shows that when we assign random PVs, both with and without errors,\footnote{When we refer to error we mean the realistic error expected from measuring PVs using SNe Ia and computed using Eq.\eqref{PVerror}.} the results are always compatible with zero. The results of this basic test are what we expect, since random PVs are not correlated, and hence have no information about the large-scale structure. We also run the test using the true PVs from the simulation, with and without errors. In both cases we are always able to recover the input value for the growth-rate parameter. The results of these tests show that the maximum likelihood always behaves as expected. Figure~\ref{fig:fs8_test_systematics} shows also the result using PVs measured from SNe Ia from Eq.\eqref{PVestimator} for comparison. 

\section{Fit results of the Hubble diagram parameters}\label{app:fitHDpara}

In this Appendix we show the fit results for the other parameters of the maximum likelihood method. We do not show the results on $\sigma_u$ since those are prior dominated, see also \citet{carreresGrowthrateMeasurementTypeIa2023} for detailed discussion about the prior on $\sigma_u$ parameter.

\begin{figure*}
    \centering
    \includegraphics[width=\hsize]{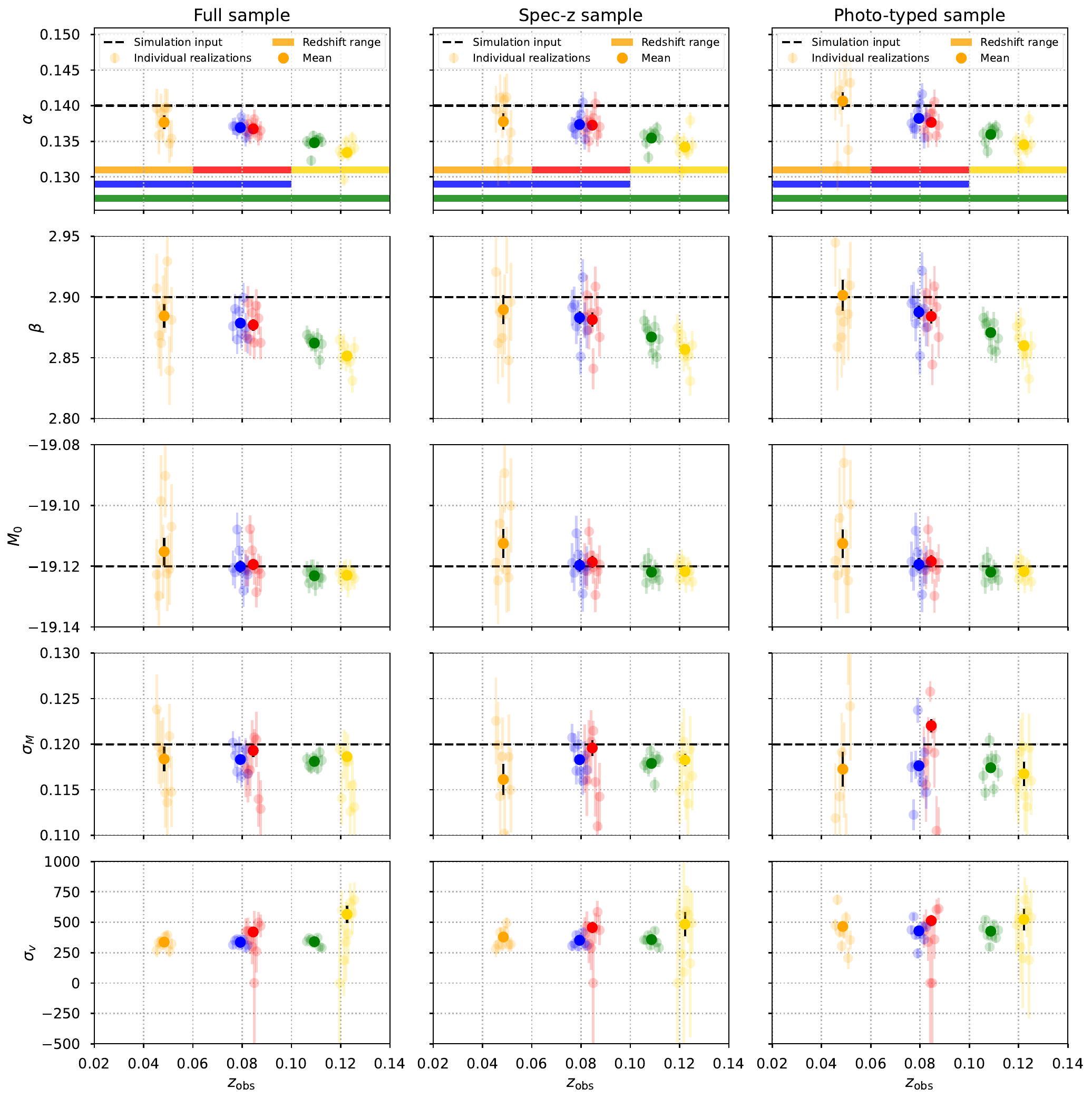}
    \caption{Result of HD parameter fits for all the scenarios in different redshift bins. \textit{Left:} Full sample, \textit{centre:} Spec-z sample, and \textit{right:} Photo-typed sample. For each scenario, the top panel shows the results for $\alpha$, then in order going down: $\beta$,$M_0$, $\sigma_M$ and $\sigma_v$. The results are shown for each of the 8 LSST realizations together with the mean value. The averages are plotted at the mean redshift positions, while the results for the individual realizations are slightly shifted along the x-axis to improve the clarity of the figure. For each scenario, the black dotted line shows the value of the simulation input. The colored bars show the redshift range used for each fit, the bars have the same colors as the measurements.}
    \label{fig:HD_shell_vol}
\end{figure*}

Figure~\ref{fig:HD_shell_vol} shows the fit results of the HD parameters and $\sigma_v$ for the Full sample, Spec-z sample and Photo-typed sample in the same redshift bins as the ones used for $f\sigma_8$ measurements.
Figure~\ref{fig:HD_shell_vol} shows that the results for $\alpha$ and $\beta$ start to be biased as the redshift increases. A bias of $\sim 0.01$ in $\alpha$ and $\sim 0.1$ in $\beta$ shifts the estimated PVs by $\sim 40$ km/s at $z=0.1$. This bias in the PVs is one order of magnitude smaller than the measurement errors at that redshift. Therefore, the bias in $\alpha$ and $\beta$ is small and has no impact on the growth-rate measurements.
The small bias in $\alpha$ and $\beta$ results indicates the presence of the Malmquist bias, as discussed in Sect.~\ref{selection_effect_HD_Vel}. Moreover, we notice that the mean $M_0$ we recover in the lowest redshift bin is slightly higher than the input value, but still compatible within the error. This is the effect of PVs and is extensively discussed in \citet{carreres2024ztfdr2}.

\newpage

\section{$f\sigma_8$ forecast across the LSST years}\label{forecast_across_year}

\begin{figure}
    \centering
    \includegraphics[width=\hsize]{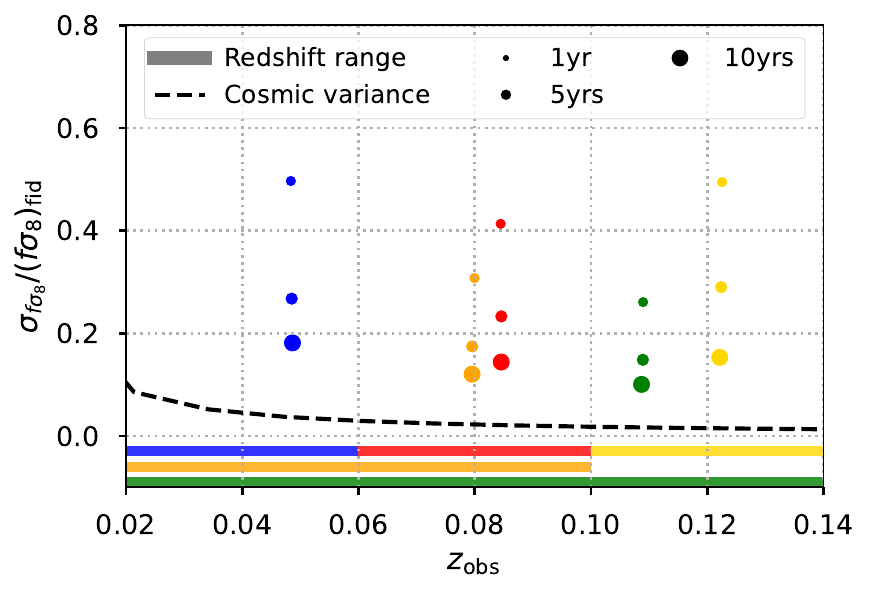}
    \caption{Error on the growth-rate parameter across the LSST duration in different redshift bins. The redshift bins are shown by the colored bars at the bottom of the figure. The points in the same redshift bin have the same color, but different sizes depending on the years. The dotted black line shows the cosmic variance limit.}
    \label{fig:fs8_SNN_DR}
\end{figure}

Using the Photo-typed sample as our baseline, we produce forecasts for the expected error on the growth rate across the LSST survey years of observation. Figure~\ref{fig:fs8_SNN_DR} shows the results of our forecast on the $f\sigma_8$ uncertainty as a function of LSST years-of-observation, considering the rolling cadence, for the different redshift bins used in Sect.\ref{result}.
All the results shown in Fig. \ref{fig:fs8_SNN_DR} are the average over the 8 LSST realizations. The cosmic variance limit shown in the Figure is computed using the Fisher Matrix formalism implemented in \texttt{FLIP} \citep[see][]{ravoux2025flip}. 
To compute the Fisher forecast we assume a density one million times the one shown in Fig.~\ref{fig:red_distr_all_samples}, and we fix $\sigma_v = 300$ \kms\,, $\sigma_M=0.12$ and $\sigma_u = 15$~\hmpc\,.
Figure~\ref{fig:fs8_SNN_DR} shows that we have a great improvement on the precision passing from year 1 to year 5 reaching $\sim 20 \%$, which is similar to the constraint from ZTF SNe Ia ($19 \%$ at $z<0.06$, see \citet{carreresGrowthrateMeasurementTypeIa2023}). After 5 years of observations the precision continues to improve reaching $12\%$ in the redshift range $0.02<z<0.14$. We conclude that after $5-6$ years of LSST we will be competitive in measuring the growth-rate parameter from SN Ia PVs.

\end{appendix}


\end{document}